\newif\ifprd
\prdtrue 

\ifprd
\documentclass[aps,prd,twocolumn,superscriptaddress,showpacs,nofootinbib]{revtex4}
\else
\documentclass[12pt,twoside,letterpaper,doublespace]{article}
\newenvironment{acknowledgments}{\section*{Acknowledgements}}{}
\fi
\newenvironment{prdappendix}{\section*{APPENDIX}}{}

\usepackage{graphicx}
\usepackage{dcolumn}
\usepackage{bm}
\usepackage{lineno}
\usepackage{setspace}
\usepackage{amssymb}
\usepackage{xspace}
\usepackage{amsmath}
\usepackage{amsthm}
\usepackage{multirow}
\ifprd
\usepackage[hyperref]{xcolor}
\usepackage{doi,hyperref}
\definecolor{darkblue}{cmyk}{0.9921,0.9708,0.0341,0.005}
\hypersetup{colorlinks,urlcolor=darkblue,citecolor=darkblue}
\else
\usepackage[hyperref]{xcolor}
\usepackage{doi,hyperref}
\definecolor{darkblue}{cmyk}{0.9921,0.9708,0.0341,0.005}
\hypersetup{colorlinks,urlcolor=darkblue,citecolor=darkblue}
\fi

\newcommand{\ppbar}{\ensuremath{p\bar{p}}\xspace}
\newcommand{\Wboson}{\ensuremath{W^{\pm}}\xspace}
\newcommand{\zzero}{\ensuremath{Z}\xspace}

\newcommand{\Zboson}{\zzero}
\newcommand{\zzerozzero}{\ensuremath{\Zboson \Zboson}\xspace}
\newcommand{\Zprime}{\ensuremath{\Zboson^{\prime}}\xspace}
\newcommand{\PM}{\ensuremath{\pm}\xspace}
\newcommand{\ZZ}{\zzerozzero}
\newcommand{\WZ}{\ensuremath{\Wboson \zzero}\xspace}
\newcommand{\ZEE}{\ensuremath{\zzero \rightarrow ee}\xspace}
\newcommand{\ZMM}{\ensuremath{\zzero \rightarrow \mu\mu}\xspace}

\newcommand{\dy}{\ensuremath{\gamma^{*}/\Zboson}\xspace}

\newcommand{\ppdyEE}{\ensuremath{\ppbar \rightarrow \dy \rightarrow ee}\xspace}
\newcommand{\ppdyMM}{\ensuremath{\ppbar \rightarrow \dy \rightarrow \mu\mu}\xspace}
\newcommand{\XZZEE}{\ensuremath{\Xboson \rightarrow \ZZ \rightarrow eeee}\xspace}
\newcommand{\dil}{\ensuremath{\ell \ell}}
\newcommand{\WZQL}{\ensuremath{\Wboson \zzero \rightarrow jj\dil}\xspace}
\newcommand{\ZZLL}{\ensuremath{\zzero \zzero \rightarrow \dil\dil}\xspace}
\newcommand{\ZZLJ}{\ensuremath{\zzero \zzero \rightarrow \dil jj}\xspace}
\newcommand{\ZZJL}{\ensuremath{\zzero \zzero \rightarrow jj\dil}\xspace}
\newcommand{\EEEE}{\ensuremath{ee ee}\xspace}
\newcommand{\EEMM}{\ensuremath{ee \mu\mu}\xspace}
\newcommand{\MMEE}{\ensuremath{\mu\mu ee}\xspace}
\newcommand{\MMMM}{\ensuremath{\mu\mu \mu\mu}\xspace}
\newcommand{\ifb}{\ensuremath{\mathrm{fb}^{-1}}\xspace}

\newcommand{\pT}{\ensuremath{p_{\mathrm{T}}}\xspace}
\newcommand{\pt}{\pT}
\newcommand{\ET}{\ensuremath{E_{\mathrm{T}}}\xspace}
\newcommand{\Et}{\ET}
\newcommand{\MX}{\ensuremath{M_{X}}\xspace}

\newcommand{\cm}{\ensuremath{\mathrm{cm}}\xspace}
\newcommand{\tev}{\ensuremath{\mathrm{Te\kern -0.1em V}}\xspace}
\newcommand{\gev}{\ensuremath{\mathrm{Ge\kern -0.1em V}}\xspace}
\newcommand{\mev}{\ensuremath{\mathrm{Me\kern -0.1em V}}\xspace}
\newcommand{\kev}{\ensuremath{\mathrm{ke\kern -0.1em V}}\xspace}
\newcommand{\ev}{\ensuremath{\mathrm{e\kern -0.1em V}}\xspace}
\newcommand{\gevc}{\ensuremath{{\mathrm{Ge\kern -0.1em V\!/}c}}\xspace}
\newcommand{\mevc}{\ensuremath{{\mathrm{Me\kern -0.1em V\!/}c}}\xspace}
\newcommand{\gevcc}{\ensuremath{{\mathrm{Ge\kern -0.1em V\!/}c^2}}\xspace}
\newcommand{\tevcc}{\ensuremath{{\mathrm{Te\kern -0.1em V\!/}c^2}}\xspace}
\newcommand{\mevcc}{\ensuremath{{\mathrm{Me\kern -0.1em V\!/}c^2}}\xspace}
\newcommand{\roots}{\ensuremath{\sqrt{s}}\xspace}

\newcommand{\cdfii}{CDF~II\xspace}

\newcommand{\abseta}{\ensuremath{|\eta|}\xspace}

\def\pythia      {\mbox{\sc pythia}\xspace}

\def\herwig      {\mbox{\sc herwig}\xspace}

\newcommand{\Xboson}{\ensuremath{X}\xspace}
\newcommand{\fb}{\ensuremath{{\rm fb}}\xspace}
\newcommand{\pb}{\ensuremath{{\rm pb}}\xspace}
\newcommand{\lumfb}{\ensuremath{\fb^{-1}}\xspace}
\newcommand{\lumpb}{\ensuremath{\pb^{-1}}\xspace}
\newcommand{\TeV}{\tev}
\newcommand{\GeV}{\gev}
\newcommand{\ndof}{\ensuremath{n_{\text{dof}}}\xspace}

\newcommand{\XZZ}{\ensuremath{\Xboson \rightarrow \zzerozzero}\xspace}
\newcommand{\ZZMassChisq}{\ensuremath{{\chi^2}_{\ZZ}}\xspace}

\begin{document}

\ifprd
\else
\linenumbers
\doublespace
\fi

\affiliation{Institute of Physics, Academia Sinica, Taipei, Taiwan 11529, Republic of China} 
\affiliation{Argonne National Laboratory, Argonne, Illinois 60439, USA} 
\affiliation{University of Athens, 157 71 Athens, Greece} 
\affiliation{Institut de Fisica d'Altes Energies, ICREA, Universitat Autonoma de Barcelona, E-08193, Bellaterra (Barcelona), Spain} 
\affiliation{Baylor University, Waco, Texas 76798, USA} 
\affiliation{Istituto Nazionale di Fisica Nucleare Bologna, $^z$University of Bologna, I-40127 Bologna, Italy} 
\affiliation{University of California, Davis, Davis, California 95616, USA} 
\affiliation{University of California, Los Angeles, Los Angeles, California 90024, USA} 
\affiliation{Instituto de Fisica de Cantabria, CSIC-University of Cantabria, 39005 Santander, Spain} 
\affiliation{Carnegie Mellon University, Pittsburgh, Pennsylvania 15213, USA} 
\affiliation{Enrico Fermi Institute, University of Chicago, Chicago, Illinois 60637, USA}
\affiliation{Comenius University, 842 48 Bratislava, Slovakia; Institute of Experimental Physics, 040 01 Kosice, Slovakia} 
\affiliation{Joint Institute for Nuclear Research, RU-141980 Dubna, Russia} 
\affiliation{Duke University, Durham, North Carolina 27708, USA} 
\affiliation{Fermi National Accelerator Laboratory, Batavia, Illinois 60510, USA} 
\affiliation{University of Florida, Gainesville, Florida 32611, USA} 
\affiliation{Laboratori Nazionali di Frascati, Istituto Nazionale di Fisica Nucleare, I-00044 Frascati, Italy} 
\affiliation{University of Geneva, CH-1211 Geneva 4, Switzerland} 
\affiliation{Glasgow University, Glasgow G12 8QQ, United Kingdom} 
\affiliation{Harvard University, Cambridge, Massachusetts 02138, USA} 
\affiliation{Division of High Energy Physics, Department of Physics, University of Helsinki and Helsinki Institute of Physics, FIN-00014, Helsinki, Finland} 
\affiliation{University of Illinois, Urbana, Illinois 61801, USA} 
\affiliation{The Johns Hopkins University, Baltimore, Maryland 21218, USA} 
\affiliation{Institut f\"{u}r Experimentelle Kernphysik, Karlsruhe Institute of Technology, D-76131 Karlsruhe, Germany} 
\affiliation{Center for High Energy Physics: Kyungpook National University, Daegu 702-701, Korea; Seoul National University, Seoul 151-742, Korea; Sungkyunkwan University, Suwon 440-746, Korea; Korea Institute of Science and Technology Information, Daejeon 305-806, Korea; Chonnam National University, Gwangju 500-757, Korea; Chonbuk National University, Jeonju 561-756, Korea} 
\affiliation{Ernest Orlando Lawrence Berkeley National Laboratory, Berkeley, California 94720, USA} 
\affiliation{University of Liverpool, Liverpool L69 7ZE, United Kingdom} 
\affiliation{University College London, London WC1E 6BT, United Kingdom} 
\affiliation{Centro de Investigaciones Energeticas Medioambientales y Tecnologicas, E-28040 Madrid, Spain} 
\affiliation{Massachusetts Institute of Technology, Cambridge, Massachusetts 02139, USA} 
\affiliation{Institute of Particle Physics: McGill University, Montr\'{e}al, Qu\'{e}bec, Canada H3A~2T8; Simon Fraser University, Burnaby, British Columbia, Canada V5A~1S6; University of Toronto, Toronto, Ontario, Canada M5S~1A7; and TRIUMF, Vancouver, British Columbia, Canada V6T~2A3} 
\affiliation{University of Michigan, Ann Arbor, Michigan 48109, USA} 
\affiliation{Michigan State University, East Lansing, Michigan 48824, USA}
\affiliation{Institution for Theoretical and Experimental Physics, ITEP, Moscow 117259, Russia}
\affiliation{University of New Mexico, Albuquerque, New Mexico 87131, USA} 
\affiliation{Northwestern University, Evanston, Illinois 60208, USA} 
\affiliation{The Ohio State University, Columbus, Ohio 43210, USA} 
\affiliation{Okayama University, Okayama 700-8530, Japan} 
\affiliation{Osaka City University, Osaka 588, Japan} 
\affiliation{University of Oxford, Oxford OX1 3RH, United Kingdom} 
\affiliation{Istituto Nazionale di Fisica Nucleare, Sezione di Padova-Trento, $^{aa}$University of Padova, I-35131 Padova, Italy} 
\affiliation{LPNHE, Universite Pierre et Marie Curie/IN2P3-CNRS, UMR7585, Paris, F-75252 France} 
\affiliation{University of Pennsylvania, Philadelphia, Pennsylvania 19104, USA}
\affiliation{Istituto Nazionale di Fisica Nucleare Pisa, $^{bb}$University of Pisa, $^{cc}$University of Siena and $^{dd}$Scuola Normale Superiore, I-56127 Pisa, Italy} 
\affiliation{University of Pittsburgh, Pittsburgh, Pennsylvania 15260, USA} 
\affiliation{Purdue University, West Lafayette, Indiana 47907, USA} 
\affiliation{University of Rochester, Rochester, New York 14627, USA} 
\affiliation{The Rockefeller University, New York, New York 10065, USA} 
\affiliation{Istituto Nazionale di Fisica Nucleare, Sezione di Roma 1, $^{ee}$Sapienza Universit\`{a} di Roma, I-00185 Roma, Italy} 

\affiliation{Rutgers University, Piscataway, New Jersey 08855, USA} 
\affiliation{Texas A\&M University, College Station, Texas 77843, USA} 
\affiliation{Istituto Nazionale di Fisica Nucleare Trieste/Udine, I-34100 Trieste, $^{ff}$University of Trieste/Udine, I-33100 Udine, Italy} 
\affiliation{University of Tsukuba, Tsukuba, Ibaraki 305, Japan} 
\affiliation{Tufts University, Medford, Massachusetts 02155, USA} 
\affiliation{University of Virginia, Charlottesville, VA  22906, USA}
\affiliation{Waseda University, Tokyo 169, Japan} 
\affiliation{Wayne State University, Detroit, Michigan 48201, USA} 
\affiliation{University of Wisconsin, Madison, Wisconsin 53706, USA} 
\affiliation{Yale University, New Haven, Connecticut 06520, USA} 
\author{T.~Aaltonen}
\affiliation{Division of High Energy Physics, Department of Physics, University of Helsinki and Helsinki Institute of Physics, FIN-00014, Helsinki, Finland}
\author{B.~\'{A}lvarez~Gonz\'{a}lez$^v$}
\affiliation{Instituto de Fisica de Cantabria, CSIC-University of Cantabria, 39005 Santander, Spain}
\author{S.~Amerio}
\affiliation{Istituto Nazionale di Fisica Nucleare, Sezione di Padova-Trento, $^{aa}$University of Padova, I-35131 Padova, Italy} 

\author{D.~Amidei}
\affiliation{University of Michigan, Ann Arbor, Michigan 48109, USA}
\author{A.~Anastassov}
\affiliation{Northwestern University, Evanston, Illinois 60208, USA}
\author{A.~Annovi}
\affiliation{Laboratori Nazionali di Frascati, Istituto Nazionale di Fisica Nucleare, I-00044 Frascati, Italy}
\author{J.~Antos}
\affiliation{Comenius University, 842 48 Bratislava, Slovakia; Institute of Experimental Physics, 040 01 Kosice, Slovakia}
\author{G.~Apollinari}
\affiliation{Fermi National Accelerator Laboratory, Batavia, Illinois 60510, USA}
\author{J.A.~Appel}
\affiliation{Fermi National Accelerator Laboratory, Batavia, Illinois 60510, USA}
\author{A.~Apresyan}
\affiliation{Purdue University, West Lafayette, Indiana 47907, USA}
% ...
\collaboration{CDF Collaboration\footnote{With visitors from $^a$University of Massachusetts Amherst, Amherst, Massachusetts 01003,
$^b$Istituto Nazionale di Fisica Nucleare, Sezione di Cagliari, 09042 Monserrato (Cagliari), Italy,
$^c$University of California Irvine, Irvine, CA  92697, 
$^d$University of California Santa Barbara, Santa Barbara, CA 93106
$^e$University of California Santa Cruz, Santa Cruz, CA  95064,
$^f$CERN,CH-1211 Geneva, Switzerland,
$^g$Cornell University, Ithaca, NY  14853, 
$^h$University of Cyprus, Nicosia CY-1678, Cyprus, 
$^i$University College Dublin, Dublin 4, Ireland,
$^j$University of Fukui, Fukui City, Fukui Prefecture, Japan 910-0017,
$^k$Universidad Iberoamericana, Mexico D.F., Mexico,
$^l$Iowa State University, Ames, IA  50011,
$^m$University of Iowa, Iowa City, IA  52242,
$^n$Kinki University, Higashi-Osaka City, Japan 577-8502,
$^o$Kansas State University, Manhattan, KS 66506,
$^p$University of Manchester, Manchester M13 9PL, England,
$^q$Queen Mary, University of London, London, E1 4NS, England,
$^r$Muons, Inc., Batavia, IL 60510,
$^s$Nagasaki Institute of Applied Science, Nagasaki, Japan, 
$^t$National Research Nuclear University, Moscow, Russia,
$^u$University of Notre Dame, Notre Dame, IN 46556,
$^v$Universidad de Oviedo, E-33007 Oviedo, Spain, 
$^w$Texas Tech University, Lubbock, TX  79609, 
$^x$Universidad Tecnica Federico Santa Maria, 110v Valparaiso, Chile,
$^y$Yarmouk University, Irbid 211-63, Jordan,
$^{gg}$On leave from J.~Stefan Institute, Ljubljana, Slovenia, 
}}
\noaffiliation

\title{Search for New Heavy Particles Decaying to $\zzerozzero \to \dil\dil, \dil jj$ in \ppbar Collisions at \roots = 1.96 \tev}
\vspace*{2.0cm}

\begin{abstract}
We report on a search for anomalous production of \Zboson boson pairs
through a massive resonance decay in data corresponding to
2.5--2.9 \lumfb of integrated luminosity in \ensuremath{p \bar{p}}
collisions at \ensuremath{\sqrt{s} = 1.96\;\TeV} using the \cdfii
detector at the Fermilab Tevatron. This analysis, with more data and
channels where the \Zboson bosons decay to muons or jets, supersedes
the 1.1 \lumfb four-electron channel result previously published by
CDF. In order to maintain high efficiency for muons, we use a new
forward tracking algorithm and muon identification requirements
optimized for these high signal-to-background channels. Predicting the
dominant backgrounds in each channel entirely from sideband data
samples, we observe four-body invariant mass spectra above
$300\,\gevcc$ that are consistent with background. We set limits using
the acceptance for a massive graviton resonance that are 7--20 times
stronger than the previously published direct limits on resonant
\ZZ diboson production.
\end{abstract}

\ifprd
\fi

\ifprd
\maketitle
\else
\newpage
\fi

\section{\label{sec:Intro}Introduction}

The standard model of particle physics (SM) has been enormously
successful, but many key questions remain to be answered by a more
complete theory. New theoretical ideas can be tested with collider
experiments, but it is also worthwhile for experiments to search
broadly for anomalous ``signatures''. One common tactic is to look
where experiments are keenly sensitive. The consummate example of this
method is the \Zprime boson search, in which a low-background,
well-understood observable (the dilepton invariant mass) is used to
constrain the new physics models that predict a dilepton
resonance. Diboson resonance searches are an attractive analog of
the \Zprime boson searches, involving higher multiplicities of the
same outgoing particles and additional mass constraints, both of which
serve to further suppress experimental backgrounds. Dibosons are the
dominant channels for high mass higgs searches, and new physics
scenarios predict particles such as Randall-Sundrum gravitons which
would decay into dibosons \cite{Randall:1999p89}. The irreducible SM
diboson background processes occur at such a low rate that they have
only recently been observed at the
Tevatron \cite{CDFCollaboration:2008p575,D0Collaboration:2008p574} at
low diboson mass ($M_{\ZZ} < 300\;\gevcc$). At high diboson mass
($M_{\ZZ} > 300\;\gevcc$) the total backgrounds are tiny.

This article presents a search for a diboson resonance in data
corresponding to 2.5--2.9 \lumfb of integrated luminosity in
$\sqrt{s}=1.96$ TeV $p \bar{p}$ collisions at the \cdfii detector at
the Fermilab Tevatron, in the decay channel $X\rightarrow
ZZ$. These \ZZ diboson processes have been well-studied at the LEP
experiments, which observed no significant deviation from the standard model
expectation up to an $e^+ e^-$ center-of-mass energy of
$207\,\GeV$ \cite{ALEPHCollaboration:1999p2245,DELPHICollaboration:2003p2242,L3Collaboration:2003p2243,OPALCollaboration:2003p2244}. The
LEP data place only indirect constraints on heavier, resonant \ZZ
diboson production \cite{Davoudiasl:2000p583}, however, and direct production
constraints at high \ZZ diboson masses\ must be probed at hadron colliders. To
the previously published CDF search for four-electron production
via \XZZEE with data corresponding to 1.1 \lumfb of integrated
luminosity \cite{CDFCollaboration:2008p1237}, we now add the dijet
channels $ee jj$ and $\mu\mu jj$, which improve sensitivity at very
high \Xboson masses where their background is negligible, and the
four-electron or -muon channels $e e\mu\mu$ and $\mu\mu\mu\mu$, which
contribute sensitivity to new physics at intermediate masses where
the \Zboson + jet(s) backgrounds are larger.

Because there are four or more outgoing leptons or quarks, the
analysis is sensitive to single lepton and jet reconstruction
efficiencies to approximately the fourth power. In particular, events
may often have one or more leptons with $|\eta|>1$ where muon
acceptance and tracking efficiency are lower than for
$|\eta|<1$. Consequently the four-lepton channels motivate development
and use of techniques to improve electron and muon reconstruction and
identification efficiencies while exploiting the kinematics of the
signature to keep backgrounds low. To augment forward muon coverage,
the present analysis also employs, for the first time, a new method of
reconstructing charged particles in the silicon detectors using
constraints from particle traces in forward regions with partial wire
tracker coverage.

The aim of the search is sensitivity to any massive particle that
could decay to \ZZ. Though we avoid focus on any one specific model,
we choose a benchmark process that is implemented in several popular
Monte Carlo generation programs, the virtual production of gravitons
in a simple Randall-Sundrum RS1
scenario \cite{Randall:1999p89,Randall:1999p88}, to fix acceptance for
the search and quantify its sensitivity. The geometry of the model
consists of two three-branes separated from each other by a single
extra dimension. Boundary conditions at the branes quantize the
momentum in the extra dimension, leading to a Kaluza-Klein tower of
discrete, massive gravitons. In RS1 scenarios with the standard model
particles confined to either brane, a discovery would involve graviton
decays directly to photons or leptons---the graviton branching ratio
to \ZZ is significant but the \Zboson boson branching ratio to leptons is
small. In more complex but well-motivated scenarios with SM particles
allowed in the extra dimension, however, graviton decays to photons,
leptons, and light jets can be suppressed, and dibosons become an
important discovery
channel \cite{Agashe:2007p1738,Fitzpatrick:2007p626}.

The organization of this article is as follows: Section 2 describes
relevant components of the \cdfii detector; Section 3, the event
selection; Section 4, the background estimates; and Section 5, the
results.

\section{The \cdfii Detector\label{sec:cdf}}
The \cdfii detector is a general purpose magnetic spectrometer
surrounded by electromagnetic and hadronic calorimeters and muon
detectors designed to record Tevatron \ppbar collisions. We briefly
describe the components of the detector relevant to this search. A
complete description can be found
elsewhere \cite{CDFCollaboration:2007p571}.

A combination of tracking systems reconstructs the trajectories and
measures momenta of charged particles in a 1.4 T solenoidal magnetic
field. Trajectories of charged particles are reconstructed using an
eight-layer silicon microstrip vertex tracker \cite{Sill:2000p594} at
radii $1.3 < r < 29$ cm from the nominal beamline\footnote{CDF uses a
cylindrical coordinate system in which $\theta$ ($\phi$) is the polar
(azimuthal) angle, $r$ is the radius from the nominal beam axis, and
$+z$ points along the proton beam direction and is zero at the center
of the detector.  The pseudorapidity is defined as $\eta \equiv -\ln
\tan(\theta/2)$.  Energy (momentum) transverse to the beam is defined $\Et \equiv E \sin
\theta$ ($\pt \equiv p \sin \theta$), where $E$ is energy and $p$ is momentum.}
and a 96-layer open-cell drift chamber (COT) providing eight superlayers of alternating stereo
and axial position measurements \cite{Affolder:2004p598} at large radii
$43 < r < 132$~cm. The COT provides full geometric coverage for
$\abseta < 1.0$. The average radius of the lowest radius axial (stereo)
COT superlayer is 58 (46) \cm, providing partial coverage for $\abseta
< 1.7 (1.9)$. The silicon
tracker provides full coverage for $\abseta < 1.8$.

Outside the tracking volume, segmented electromagnetic (EM)
lead-scintillator and hadronic (HAD) iron-scintillator sampling
calorimeters measure particle energies \cite{Balka:1988p588}.  The
central ($\abseta < 1.1$) calorimeters are arranged around the
interaction point in a projective-tower cylindrical geometry, divided
azimuthally into $15^{\circ}$ wedges. This calorimeter measures
electron energies with a resolution of $[\sigma(E)/E]^2
=(13.5\%)^2/E_T + (2\%)^2$.  The forward calorimeters ($1.1 < \abseta
< 3.6$) are arranged in an azimuthally-symmetric disk geometry and
measure electron energies with a resolution of $[\sigma(E)/E]^2 =
(16.0\%)^2/E + (1\%)^2$.  Wire chambers (scintillator strips) embedded
in the central (forward) EM calorimeters at $\sim 6 X_0$, the average
depth of shower maximum, provide position and lateral shower
development measurements for $\abseta < 2.5$.

Beyond the calorimeters, muon drift chambers and scintillators measure
particles that traverse the entire inner and outer detectors and
reject the instrumental backgrounds of the central muon triggers. The
central muon chambers (CMU) lie just outside the central hadronic
calorimeter with $\phi$-dependent coverage for $0.03
< \abseta < 0.63$. The central muon upgrade (CMP) augments the CMU
coverage in $\phi$ and lies behind another approximately 3 interaction
lengths of steel. The central muon extension (CMX) extends coverage
into the region $0.65 < \abseta < 1.0$.

The beam luminosity is determined by measuring the inelastic \ppbar
collision rate with gas Cherenkov detectors \cite{Acosta:2001p592},
located in the region $3.7 < \abseta < 4.7$.

At each bunch crossing, a three-level trigger
system \cite{CDFCollaboration:2007p571} scans the detector output for
$\abseta < 1.1$ electrons or $\abseta < 1.0$ muons with at least
18 \gevc of transverse momentum. We accept events that satisfy one of
four trigger paths: one that requires a deposition of at least 18 \gev
transverse energy in the calorimeter consistent with an electron and a matching
COT track with at least 9 \gevc of transverse momentum; another with
fewer electron identification requirements intended to ensure high
efficiency for very energetic electrons; a muon path requiring a COT
track with at least 18 \gevc of transverse momentum pointing toward
signals in both the CMU and CMP chambers (a CMUP trigger) and
traversing the calorimeter consistent with a minimum-ionizing
particle; or a similar muon path with signals in the CMX chamber
instead of the CMU and CMP chambers.

\section{Data Collection}\label{sec:collection}

We use data corresponding to an integrated luminosity of 2.5--2.9~\ifb
depending on the data quality criteria applicable to the relevant \ZZ
diboson decay channel. We separately analyze six channels: $ee ee$,
$ee \mu\mu$, $\mu\mu ee$, $\mu\mu \mu\mu$, $ee jj$, and $\mu\mu
jj$. Events are divided into the six categories based on the trigger,
where the first lepton denotes the required trigger path, and the
presence of lepton and jet candidates identified using the criteria
listed in Tables \ref{table:cem} through \ref{table:jets}. The trigger
lepton criteria are the most stringent; subsequent kinematic signature
selections yield very low backgrounds, allowing very efficient
identification criteria to be used for the other lepton
candidates. Events accepted by either electron trigger path and
containing at least one electron candidate that fired the electron
trigger and satisfied the offline selection criteria are excluded from
the muon-triggered categories. There are no events that satisfy the
requirements of more than one category.

During this selection we identify the events containing at least two
leptons (including the trigger lepton) using the nominal CDF event
reconstruction software. We reprocess these events using a revision of
the software with improved tracking, including more efficient forward
tracking algorithms, and then select all final particles from the
reprocessed data. In this way, we avoid CPU-intensive reconstruction
of a 96.6\% subset of the sample that has no chance to pass our
final selection. Nevertheless, two subsets of the data corresponding
to $200\:\lumpb$ of integrated luminosity each were fully reprocessed
without the initial two-lepton selection and analyzed to confirm that
the procedure used for the full dataset is fully efficient for events
of interest to the analysis.

\begin{table}
  \caption{Calorimeter electron identification criteria. We require
  Had/EM, the ratio of energies measured in the hadronic and
  electromagnetic calorimeters, to be less than $f(E) = 0.055 +
  0.00045 \times (E / \GeV)$ where $E$ is the measured calorimeter
  energy. $\text{Iso}_{\text{cal}}$ is the calorimeter energy
  measured within $\Delta R = 0.4$ centered on the electron, excluding the
  electron energy. $\eta_{\mathrm{det}}$ is calculated assuming an
  origin at $z=0$. LshrTrk is a lateral shower shape variable
  described in Ref. \cite{Abe:1991p1721}.}
  \label{table:cem}
  \begin{center}
    \begin{tabular}{cccc}
      { Criteria}            & { Trigger} & { Central} & { Forward}\\ \hline \hline
      \noalign{\vspace{1mm}} 
      $E_T$(\gev)		  & $> 20 $ & $>5$    & $>5$ \\
      $|$Track $z_0|$ (cm)& $< 60 $ & $<60$   & \\
      Had/EM       		  & $< f(E)$ & $< f(E)$ & $< 0.05$ \\
      $\text{Iso}_{\text{cal}}$/$E_{\text{cal}}$       & $< 0.2 $ & $< 0.2$ & $<0.2$ \\
      $|\eta_{\mathrm{det}}|$ & &  & $< 2.5$ \\
      LshrTrk 		      & $< 0.4 $ &  & \\
      Track $p_T$ (\gevc) & $> 10 $ &  & \\ \hline \hline
    \end{tabular}
  \end{center}
\end{table}

\begin{table}
  \caption{Track electron identification
  criteria. Tracks must consist of measurements in several COT superlayers. Silicon measurements are not required. $\text{Iso}_{\text{trk}}$ is the scalar sum of the momenta of all
  tracks measured within a circle of $\Delta R = 0.4$ centered on the electron track
  direction. $\Delta R_{EM}$ is the separation in the $\eta-\phi$ plane
  between the electron track and the nearest calorimeter electron cluster, as defined in Table \ref{table:cem}.}
  \label{table:trackem}
  \begin{center}
    \begin{tabular}{cc}
      { Criteria} &   \\ \hline \hline
      \noalign{\vspace{1mm}} 
      $p_T\,$(\gevc)  		& $> 10$  \\
      Axial Superlayers 	& $> 3$ \\
      Stereo Superlayers 	& $> 2 $ \\
      $|$Track $z_0|$ (cm)	& $< 60 $ \\
      $p_{\text{trk}}$ / $(\text{Iso}_{\text{trk}} + p_{\text{trk}})$ & $> 0.9 $\\
      $|d_0|$ & $ < \begin{cases} 
        200\:\text{$\mu$m}&\text{silicon}\\
        2\:\text{mm}&\text{no silicon}
      \end{cases}$ \\
      $\Delta R_{EM}$& $> 0.2$ \\ \hline \hline
    \end{tabular}
  \end{center}
\end{table}

The electron criteria listed in Tables \ref{table:cem}
through \ref{table:trackem} are nearly identical to the previous
$eeee$ analysis \cite{CDFCollaboration:2008p1237}. The double \Zboson boson 
mass peak signature admits little background, allowing appreciably
more efficient electron criteria than those used for many other CDF
analyses (for example, Ref. \cite{CDFCollaboration:2007p571}). We
require either an isolated calorimeter cluster with electron-like
energy deposition or, to recover acceptance, an isolated track
pointing at uninstrumented regions of the calorimeters. Such regions
constitute approximately 17\% of the solid angle for $|\eta| < 1.2$
and would otherwise reduce our four-electron acceptance by a factor of
two. The transverse energy threshold for non-trigger electrons is
5 \gev. As the mass of the signal resonance $X\rightarrow ZZ$
increases, the energies of the two \Zboson boson decay products become
asymmetric in the detector frame, and thus our criteria must
efficiently select leptons with transverse momentum of order 10 \gevc
as well as leptons with \pT of hundreds of \gevc.

\begin{table}
  \caption{Muon identification criteria. The CMU, CMP, and CMX match
  variables compare the track position extrapolated to the relevant
  muon chambers with the chamber position measurements. The
  non-trigger $p_T$ requirement is lower for tracks with muon chamber
  information attached. $\text{Iso}_{\text{cal}}$ is the sum of calorimeter
  energies measured in towers within a circle of $\Delta R = 0.4$
  centered on the muon tower. $E_{EM}$ and $E_{HAD}$ are the
  electromagnetic and hadronic calorimeter energies recorded in towers
  intersected by the muon track, and $f_{EM}(p^{trk}) = 4
  + \text{max}(0, 0.0115*(\frac{p^{trk}}{\gevc}-100))$ and
  $f_{HAD}(p^{trk}) = 12 + \text{max}(0,
  0.028*(\frac{p^{trk}}{\gevc}-100))$ are functions of the track
  momentum. The cuts on track curvature $\kappa$, its uncertainty
  $\sigma_{\kappa}$, and the $\chi^2$ probability of the fit
  $\text{Prob}(\chi^2,\ndof)$ reject poorly measured tracks.}
  \label{table:muons}
  \begin{center}
    \begin{tabular}{cccc}
      { Criteria}  & { CMUP} & { CMX} & { Non-trigger} \\ \hline \hline
      \noalign{\vspace{1mm}}
      $p_{T}^{trk} (\gevc)$ & $>20$ & $>20$ & $>2,10$\\
      CMU match & $<10\;\cm$ & & \\
      CMP match & $<20\;\cm$ & & \\
      CMX match & & $<10\;\cm$ & \\ \hline
      & \multicolumn{3}{c}{common to all categories} \\
      $\text{Iso}^{\text{cal}} / p^{trk}$ & \multicolumn{3}{c}{$< 0.2$} \\
      $E_{EM} (\gev)$ & \multicolumn{3}{c}{$< f_{EM}(p^{trk})$} \\
      $E_{HAD} (\gev)$ & \multicolumn{3}{c}{$< f_{HAD}(p^{trk})$} \\
      $\kappa / \sigma_{\kappa}$ & \multicolumn{3}{c}{$>2.5$} \\
      $\text{Prob}(\chi^2,\ndof)$ & \multicolumn{3}{c}{$>10^{-10}$} \\
      $|z_0| (cm)$ & \multicolumn{3}{c}{$< 60$ }\\
      $|d_0|$ & \multicolumn{3}{c}{$ < \begin{cases} 
        200\:\mu\text{m}  & \text{silicon} \\
        2\:\text{mm}  & \text{no silicon}
      \end{cases}$} \\ \hline \hline
    \end{tabular}
  \end{center}
\end{table}

The muon criteria listed in Table \ref{table:muons} require an
isolated track satisfying basic track quality criteria and depositing
minimal energy in the calorimeter. We make use of a new track
reconstruction algorithm, described in the Appendix, and apply less
stringent energy and isolation requirements than typical for CDF
high \pT analyses so as to increase our acceptance and efficiency. Muon
candidate tracks may be matched to information in the muon chambers,
but to recover acceptance lost due to gaps in chamber coverage and to
recover efficiency lost due to pointing requirements, chamber matching
is not required except for the trigger muon.

\begin{table}
  \caption{Jet identification criteria. The JETCLU algorithm is
  discussed in Ref. \cite{PhysRevD.45.1448}. $\Delta R_{EM}$ is the
  separation in the $\eta-\phi$ plane between centroids of the jet
  cluster and the nearest electron cluster.}
  \label{table:jets}
  \begin{center}
    \begin{tabular}{cc}
      { Selection Criteria} & \\ \hline \hline
      \noalign{\vspace{1mm}} 
      Algorithm & JETCLU 0.4 Cone \\
      $E_T^{\text{raw}}$(\gev)	      	& $>10$ \\
      $|\eta_{\text{centroid}}|$ & $< 3.64$ \\
      $E_{EM} / E_{tot}$ & $< 0.95$ \\
      $\Delta R_{EM}$ & $> 0.4$ \\   \hline \hline    
    \end{tabular}
  \end{center}
\end{table}

Jets must satisfy the criteria listed in Table \ref{table:jets} and
jet energies are corrected for instrumental
effects \cite{Bhatti:2006p795}. Before relying on the jet energy
measurements for the two-lepton two-jet analysis, we have verified
that these corrections balance transverse momentum in the $\Zboson
+ \text{jet(s)}$ events considered here. We choose kinematic
requirements on individual jet energies, on dijet invariant masses,
and on four-body masses involving jets so that the
systematic uncertainties on \XZZ signal acceptances and efficiencies
from mis-modeling of QCD radiation or jet reconstruction effects are
small.

\begin{figure} \centering
  \mbox{ \includegraphics[width=7.2cm]{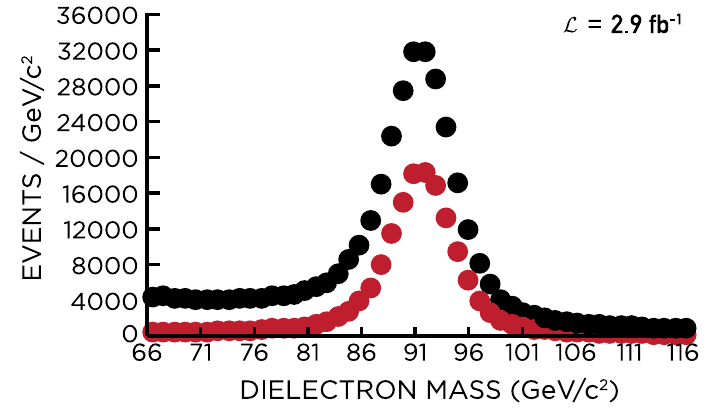}
  } \caption{$\ZEE$ yield and background comparison between dielectron
  candidates consisting of a trigger electron and an electron
  candidate satisfying either (lower set of points) the CDF standard
  electron criteria or (upper set of points) the criteria employed in
  the present analysis. The peak yield increases from about 146,000
  with the standard criteria to 256,000 candidates with our optimized
  criteria. The corresponding increase in continuum background is modest, a
  factor of 2.0 for 81--101 \gevcc. This background is later suppressed by the
  four-body kinematic selection.}\label{fig:ee_yield_comparison}
\end{figure}

\begin{figure} \centering
  \mbox{ \includegraphics[width=7.2cm]{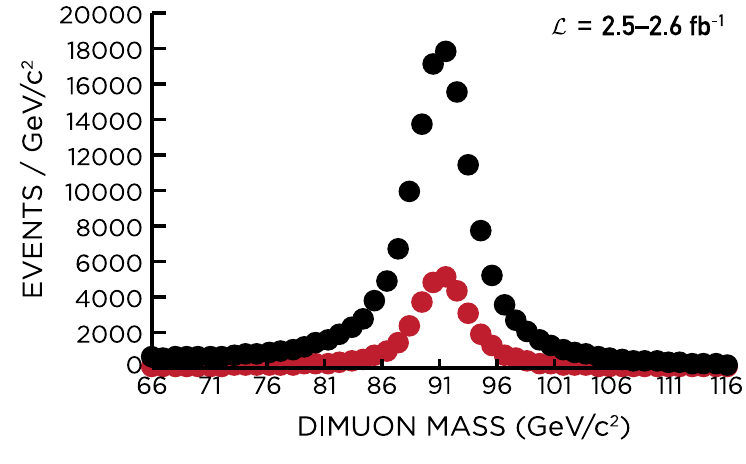}
  } \caption{$\ZMM$ yield and background comparison between dimuon
  candidates consisting of a trigger muon and an muon candidate
  satisfying either (lower set of points) the CDF standard muon
  criteria or (upper set of points) the criteria employed in the
  present analysis. The peak yield increases from about 35,000
  candidates with the standard criteria and tracking to 150,000
  candidates with our optimized criteria and tracking. The continuum
  background increases by a factor of 14 for 81--101 \gevcc. This
  background is later suppressed by the four-body kinematic
  selection.}\label{fig:mm_yield_comparison}
\end{figure}

Figs. \ref{fig:ee_yield_comparison} and
\ref{fig:mm_yield_comparison} show comparisons of the peaking and background components of the dielectron and dimuon yield. The
combination of our changes to the identification criteria and to the
tracking algorithms increases the peak yield by factors of 1.8 and
4.3, respectively.

Measurements of the \ppdyEE and \ppdyMM cross-sections provide an
important test of our understanding of the trigger, reconstruction,
and identification efficiencies and Monte Carlo modeling for this new
lepton selection. We divide the data into 18 data-taking periods and
measure each of the above efficiencies for each period. For each
period, we then compute the Drell-Yan cross-section for all
combinations of trigger and lepton type using a signal plus background
fit to the data and Drell-Yan Monte Carlo. The average instantaneous
luminosity tends to increase with data-taking period as Tevatron
upgrades were brought online. Figs. \ref{fig:ee_xsec_vs_time} and
\ref{fig:mm_xsec_vs_time} show the resultant cross-sections and their
dependence on time.

\begin{figure} \centering
  \mbox{ \includegraphics{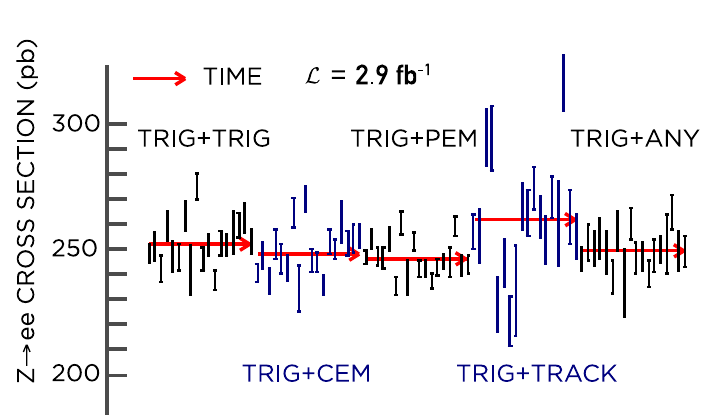} }
  \caption{\ZEE cross-sections (and averaged cross-sections) for $66 < M_{\dil} < 116\;\gevcc$ and
  various selections. The horizontal axis indicates the 18 periods
  for five selections in succession: two trigger electrons ($252.1 \pm
  1.2$ \pb), a trigger electron and a central calorimeter electron
  (TRIG+CEM, $248 \pm 1.1$ \pb), a trigger electron and a forward
  calorimeter electron (TRIG+PEM, $246.2 \pm 0.9$ \pb), a trigger
  electron and a track electron (TRIG+TRACK, $262.1 \pm 2.3$ \pb),
  and, calculated separately, the combination of a trigger electron
  and any electron selected using the analysis criteria ($249.4 \pm
  1.6$ \pb). The averaged cross sections are indicated by horizontal lines. 
  Uncertainties are statistical only with the correlated
  luminosity uncertainty not shown.}\label{fig:ee_xsec_vs_time}
\end{figure}

\begin{figure} \centering
  \mbox{ \includegraphics{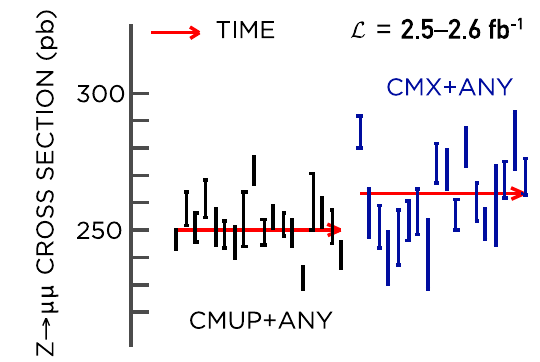} } \caption{\ZMM
  cross-sections (and averaged cross-sections) for $66 < M_{\dil} <
  116\;\gevcc$ and various selections. The horizontal axis indicates
  the 18 periods for two selections in succession: a CMUP trigger muon
  and another muon satisfying the analysis criteria ($250.0 \pm
  1.2$ \pb) and a CMX trigger muon combined with another selected muon
  ($263.4 \pm 1.8$ \pb). The averaged cross sections are indicated by
  horizontal lines.  Errors are statistical only with the correlated
  luminosity uncertainty not shown.}\label{fig:mm_xsec_vs_time}
\end{figure}

For the cases where both electron energy measurements come from the
calorimeters, the fit signal function is a Breit-Wigner fixed to the
world average \Zboson boson mass and width convolved with a Gaussian
resolution function. In the trigger plus track lepton sample, where
there is a significant radiative tail below the \Zboson boson mass, the
signal function is a Crystal Ball function. In the combined sample
both functions are included. In all cases, the background function is
an exponential. The combined result for the electron selection,
$249.4 \pm 1.6$ \pb for the mass range $66 < M_{\dil} < 116\,\gevcc$,
is in agreement with the dedicated CDF measurement
$\sigma_{\gamma*/\Zboson} \times \text{Br}(p\bar{p} \rightarrow \dy \rightarrow \dil)
= 254.9 \pm 3.3 \text{(stat.)} \pm
4.6 \text{(sys.)}$ \cite{CDFCollaboration:2007p571} and is evidence
that the constituent efficiencies and scale factors are understood
well.

Note that there are large variations in the trigger plus track
electron cross-section. In these cases, the fit to the data
underestimates or overestimates the amount of power-law background
contamination of the low dielectron mass radiative tail of
the \Zboson boson. The independent fit to the combination of all selections
has a higher signal to noise ratio and is not sensitive to this effect.

The results for the muon selection also show good agreement with the
other CDF measurements, except for a 20\% shift in the first period
and up to a 10\% shift in many of the later periods. The disrepancies
are due to imperfect modeling of tracking-related muon identification
efficiencies for the new reconstruction software. Based on these
results, we assign a 20\% systematic uncertainty on the signal
acceptance for each of the six channels. In part because much of the
data were collected with the CMUP trigger and during periods with
measured cross-sections in good agreement with expectation, and
because two of the six analysis channels do not involve muons, this
systematic uncertainty over-covers the observed variation in cross
section and is the dominant uncertainty for the
analysis. Nevertheless, the final sensitivity of the analysis improves
substantially on the earlier $eeee$
search \cite{CDFCollaboration:2008p1237}.

\section{Kinematic Analysis}

After selecting electrons, muons, and jets, we consider all possible
four-lepton $\dil\dil$ or two-lepton two-jet $\dil jj$ combinations
for each event that contains a trigger lepton. No requirement is made
on the mass or charge of dilepton pairs. Any two particles must
have a minimum separation $\Delta R = \sqrt{ {\Delta \eta}^2 +
{\Delta \phi}^2 }$ of 0.2. Dilepton pairs with tracks present for both
leptons must point back to the same $z_0$ production location in order
to suppress background from additional pileup interactions. Track
timing information is used for a very pure veto of muon and track
electron pairs consistent with cosmic rays. There are no events that
appear in more than one
\ZZ diboson decay channel.

For the four-lepton channels, we consider all possible combinations of
leptons for each event and select the one that minimizes a
$\chi^2$ variable quantifying consistency between the dilepton masses
and the \Zboson boson pole mass:
\begin{equation*}
  \chi^2_{ZZ} = {\displaystyle\sum\limits_{i=1,2}} \frac{(M_{\Zboson}^{(i)} - 91.187\,\gevcc)^2}{\sigma^2_{M^{(i)}} + \sigma_{\Gamma}^2},
\end{equation*}
where $\sigma_{M^{(i)}}$ is the detector mass resolution computed from
the individual lepton calorimeter or tracking measurements for the
dilepton mass $M_{\Zboson}^{(i)}$ and a Gaussian approximation with
$\sigma_{\Gamma} = 3.25\,\gevcc $ allows for the nonzero width of the
\Zboson boson resonance.

For the $\dil jj$ channels, we consider all possible combinations of
leptons and jets. We select the two highest \ET jets and the dilepton
pairing that minimizes the first term of the equation above. This
explicitly avoids possible \Zboson boson mass bias in the dijet mass
spectrum that would complicate the background estimate discussed in
the following section. We then require $M_{\Zboson} > 20\,\gevcc$ for
each pairing and, for the dijet channels, $\chi^2_{Z} < 25$ for the
leptonic \Zboson boson.

{\it A priori} we define our signal region to be $M_{\Xboson} >
300\,\gevcc$ so as to avoid most standard model backgrounds. For the
$\dil\dil$ modes we further require $\chi^2_{ZZ} < 50$ and for the
$\dil jj$ modes we require $65 < M_{jj} < 120\,\gevcc$. Each event may
contain additional leptons, jets, or other particles beyond the
four that contribute to the signal candidate.

\section{Background Estimates}\label{sec:background}

For both the four-lepton and the dijet channels, the dominant
backgrounds at high \MX are a mixture of \Zboson + jets, \Wboson +
jets, multi-jets, and various lower-rate processes resulting in one or more
hadrons that mimic an electron or muon. The diboson
processes \WZQL, \ZZLL, and \ZZLJ peak at $\ensuremath{\chi^2_{ZZ}} <
50$ or $65 < \ensuremath{M_{jj}} < 120\,\gevcc$, while all other
backgrounds do not peak in both \Zboson boson masses
simultaneously. We use a \pythia Monte Carlo model \cite{Pythia:2001}
with the CDF detector simulation to estimate the small contribution
from resonant diboson processes and fit sideband data to collectively
estimate all backgrounds that do not contain two bosons, collectively
referred to as non-resonant background.

We estimate the $\dil \dil$ background by extrapolating the yield in
the $185 < M_{\Xboson} < 300\,\gevcc$ region to the signal region
($M_{\Xboson} > 300\,\gevcc$ and $\chi^2_{ZZ} < 50$) using a shape
determined from a sample enhanced in non-resonant background. In order
to construct samples enriched in this background, four-lepton
candidates are selected in which some of the reconstructed leptons are
``anti-selected'' to fail one or more lepton identification
criteria. Anti-selected electrons must fail
the $\text{HAD}/\text{EM}$ energy selection and anti-selected muons must fail the
minimum-ionizing energy selection. To further increase available
statistics, the isolation requirement is removed for both
categories. Events reconstructed with the standard CDF processing
that contain at least one trigger lepton and one anti-selected lepton
are included in the reprocessing discussed in Section
\ref{sec:collection}. The $\ensuremath{\chi^2_{ZZ}}$ vs. $M_{\Xboson}$
distributions for the resultant samples are shown in Figs.
\ref{fig:x2mx_efakes} and \ref{fig:x2mx_mfakes}.

Figs. \ref{fig:etrig_fl_fakes} and \ref{fig:mutrig_fl_fakes} show
the invariant mass distributions of trigger lepton plus anti-selected
lepton pairings for electrons and muons, respectively. The absence of
an appreciable peak indicates few resonant \Zboson boson events survive the
anti-selection.

\begin{figure} \centering
  \mbox{ \includegraphics[width=7.2cm]{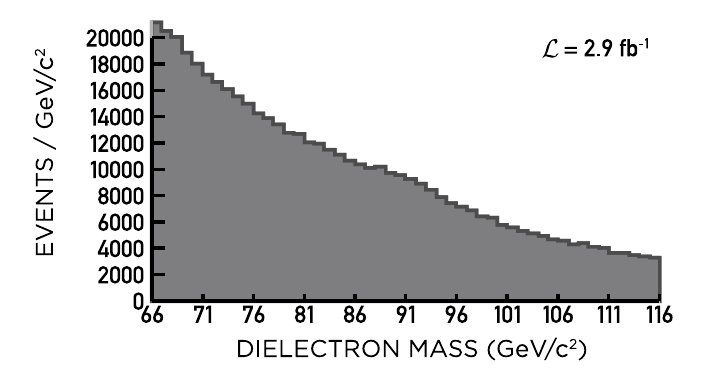}
  } \caption{Invariant mass distribution for pairs of candidates
  consisting of a trigger electron and an anti-selected
  electron.}\label{fig:etrig_fl_fakes}
\end{figure}

\begin{figure} \centering
  \mbox{ \includegraphics[width=7.2cm]{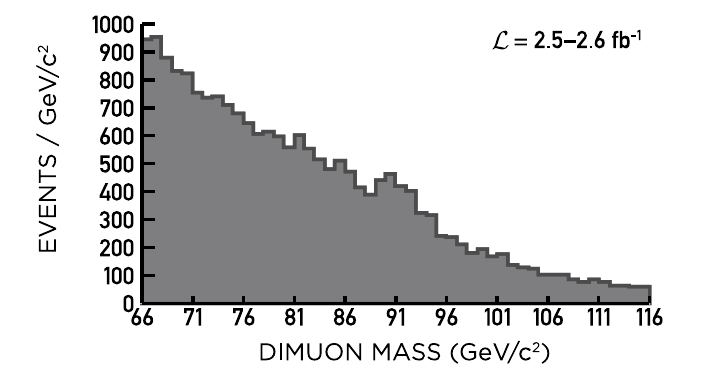}
  } \caption{Invariant mass distribution for pairs of candidates
  consisting of a trigger muon and an anti-selected
  muon.}\label{fig:mutrig_fl_fakes}
\end{figure}

\begin{figure*} \centering
  \mbox{ \includegraphics{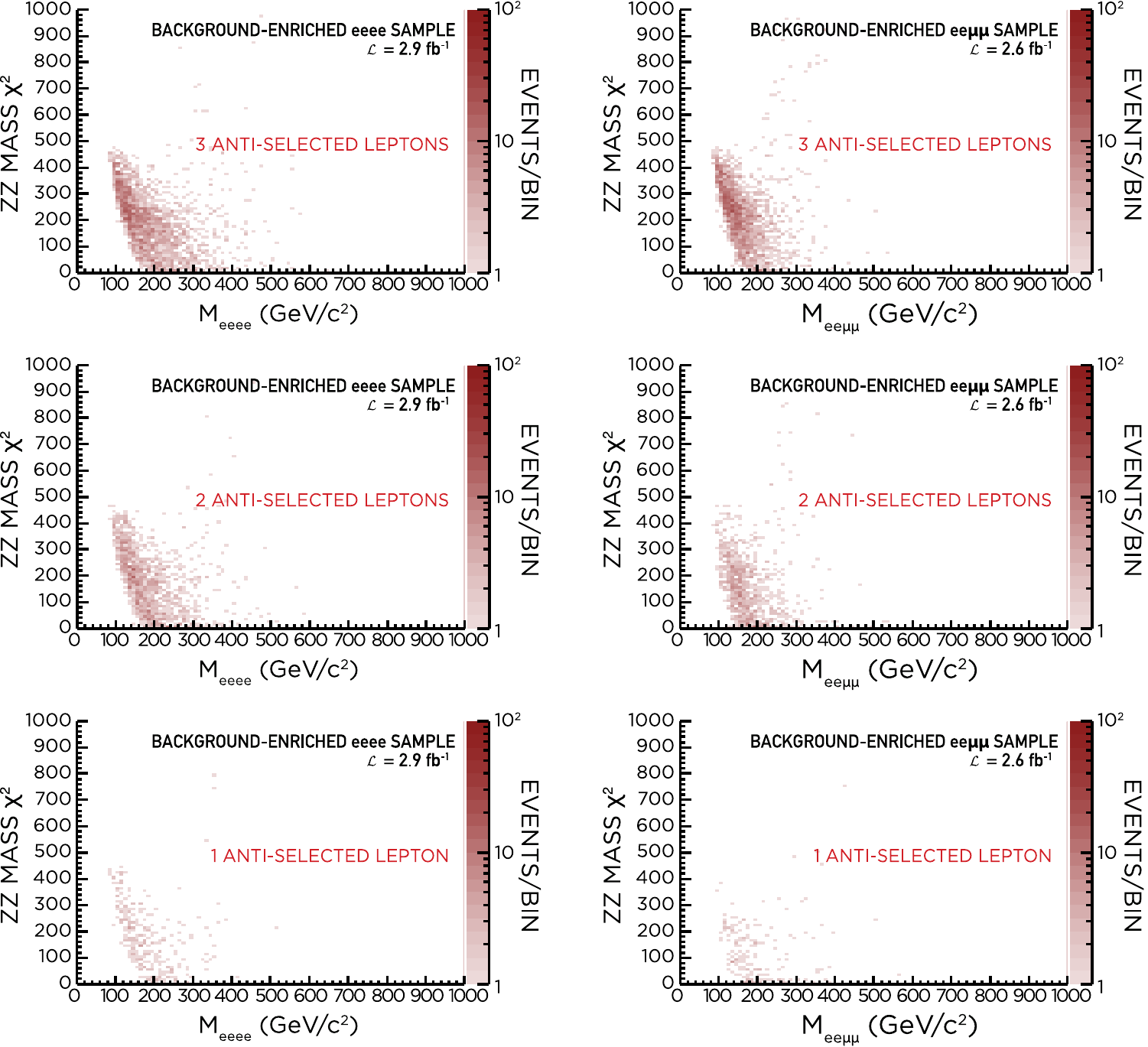} }
  \caption{$\ensuremath{\chi^2_{ZZ}}$ vs. $M_{\Xboson}$ distributions
    for the four-electron and electron-triggered two-electron two-muon
    sideband samples with 1, 2, and 3 anti-selected
    leptons.}\label{fig:x2mx_efakes}
\end{figure*}

\begin{figure*} \centering
  \mbox{ \includegraphics{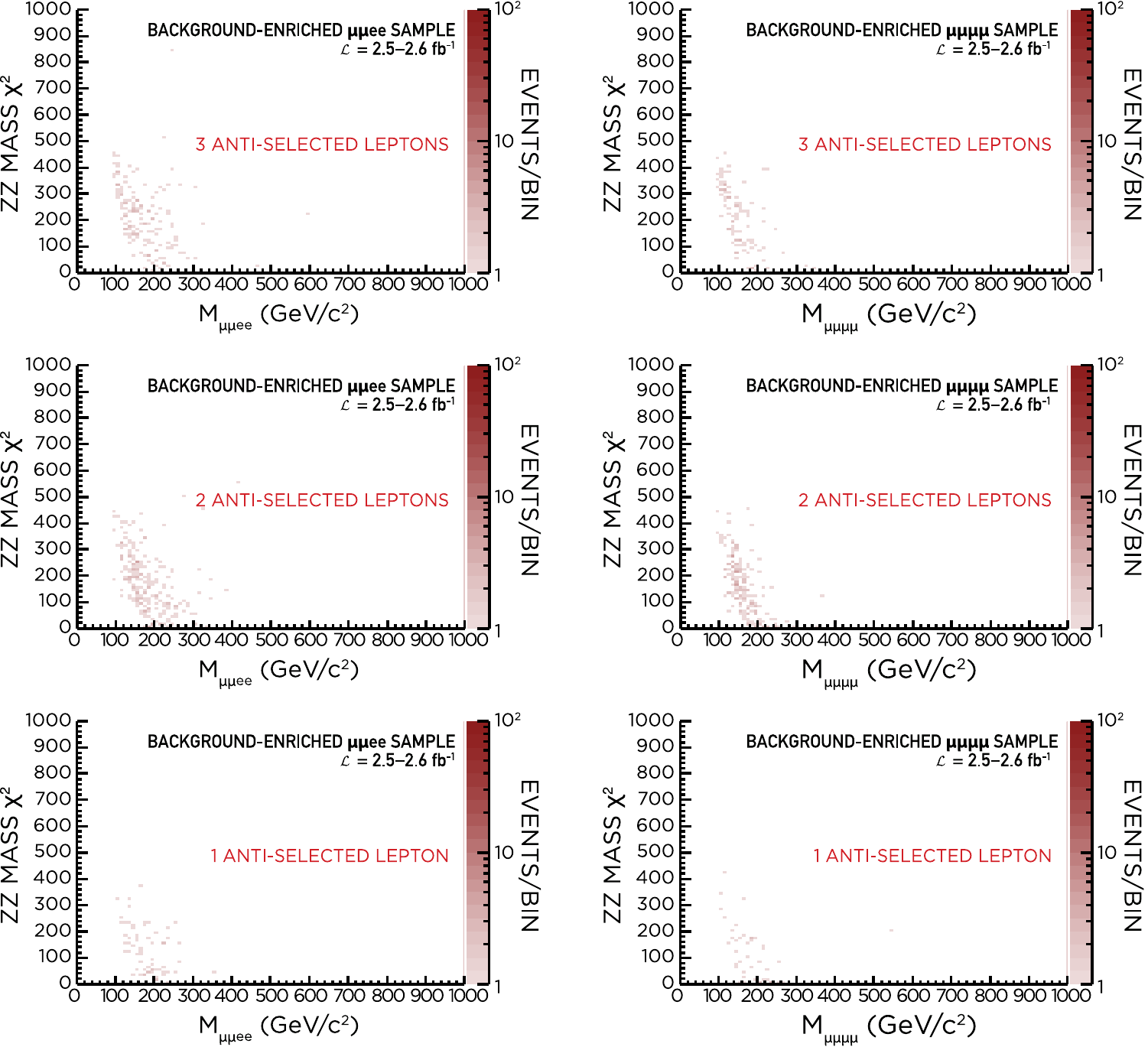} }
  \caption{$\ensuremath{\chi^2_{ZZ}}$ vs. $M_{\Xboson}$ distributions
    for the muon-triggered two-electron two-muon and four-muon
    sideband samples with 1, 2, and 3 anti-selected
    leptons.}\label{fig:x2mx_mfakes}
\end{figure*}

The two samples of four-body candidates that consist of a trigger
lepton and either two- or three-anti-selected leptons with
$M_{\Xboson} > 185\,\gevcc$ and $\chi^2_{ZZ} < 500$ are then fit
simultaneously to the empirical form \begin{equation*}
  f(\chi^2_{ZZ},M_{\Xboson}) = M_{\Xboson}^\gamma \cdot e^{\tau
    \chi^2_{ZZ}} \end{equation*} to determine the falling shape of the
$M_{\dil\dil}$ distribution (the power law parameter $\gamma$) and the
relationship of the number of events in the $\chi^2_{ZZ} < 50$ \ZZ
window to the number in the off-mass sidebands (the exponential decay
parameter $\tau$). As background composition and fake rate kinematic
dependence varies with trigger and lepton type, we fit these sidebands
separately for the \EEEE, \EEMM, \MMEE, and \MMMM background
shapes. Figs. \ref{fig:eeee_fit_projections} through
\ref{fig:mmmm_fit_projections} show one-dimensional projections of the
fit result for each channel against the fitted two- and three-anti-selected lepton
$\chi^2_{ZZ}$ and $M_{\dil\dil}$ data as well as the one
anti-selected sample, which is not used in the fit. Table \ref{table:llllfits} lists the fit
parameters obtained with their statistical uncertainty.

\begin{table}
  \caption{Four-lepton background fit results.}
  \label{table:llllfits}
  \begin{center}
    \begin{tabular}{cD..{3.2}cD..{1.2}D..{3.4}cD..{1.4}}
      \multicolumn{1}{c}{ Channel} & 
      \multicolumn{3}{c}{$\gamma$} & 
      \multicolumn{3}{c}{$\tau$} \\ \hline \hline
      \noalign{\vspace{1mm}} 
      $ee ee$ & -4.39 & \PM & 0.09 &  -0.0184  & \PM & 0.0005 \\
      $ee \mu\mu$ & -5.4 & \PM & 0.2 &  -0.0161  & \PM & 0.0005 \\
      $\mu\mu ee$ & -5.3 & \PM & 0.3 &  -0.020  & \PM & 0.002 \\
      $\mu\mu \mu\mu$ & -6.5 & \PM & 0.6 &  -0.030  & \PM & 0.003 \\ \hline \hline
    \end{tabular} 
  \end{center}
\end{table}

\begin{figure} \centering
  \mbox{ \includegraphics{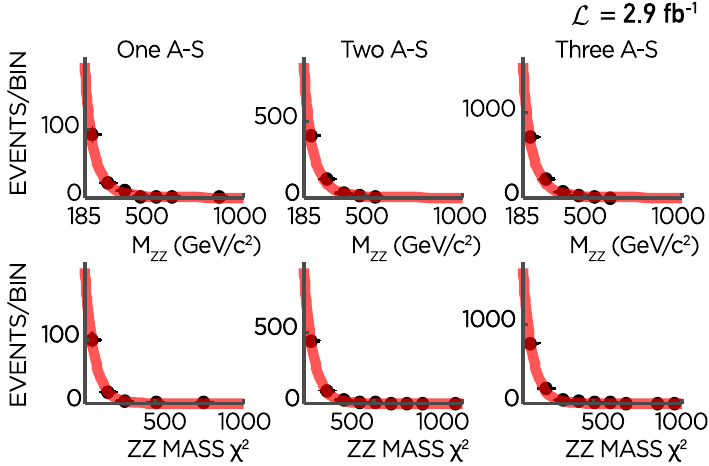}
  } \caption{$M_{eeee}$ and $\chi^2_{ZZ}$ for the one-, two-, and
  three-anti-selected (A-S) four-electron samples, and the results of the
  simultaneous non-resonant background shape fit to the two- and
  three-anti-selected electron
  samples.}\label{fig:eeee_fit_projections}
\end{figure}

\begin{figure} \centering
  \mbox{ \includegraphics{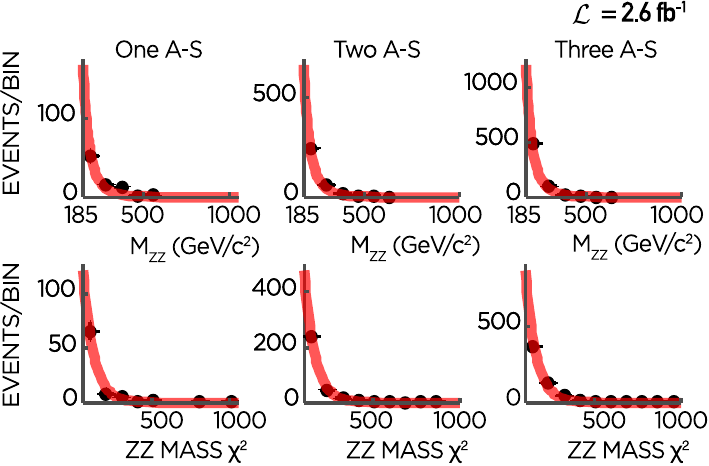} }
  \caption{$M_{ee\mu\mu}$ and $\chi^2_{ZZ}$ for the one-, two-, and
    three-anti-selected (A-S) lepton samples for the electron-triggered two-electron two-muon channel, and the results of the simultaneous non-resonant
    background shape fit to the two- and three-anti-selected lepton
    samples.}\label{fig:eemm_fit_projections}
\end{figure}

\begin{figure} \centering
  \mbox{ \includegraphics{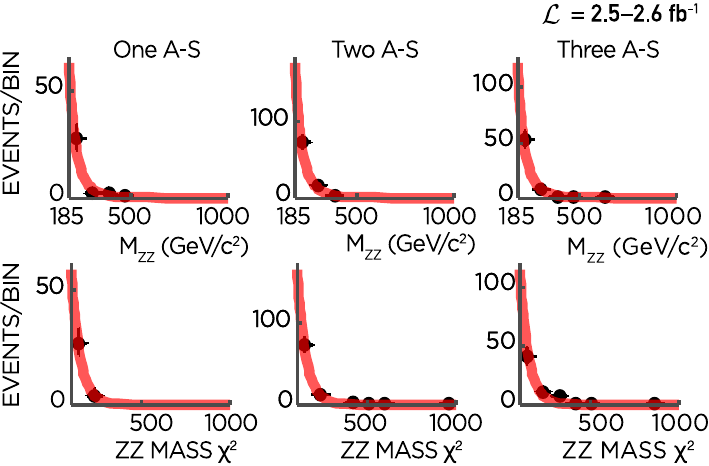} }
  \caption{$M_{\mu\mu ee}$ and $\chi^2_{ZZ}$ for the one-, two-, and
    three-anti-selected (A-S) lepton samples for the muon-triggered two-electron two-muon channel, and the results of the simultaneous non-resonant
    background shape fit to the two- and three-anti-selected lepton
    samples.}\label{fig:mmee_fit_projections}
\end{figure}

\begin{figure} \centering
  \mbox{ \includegraphics{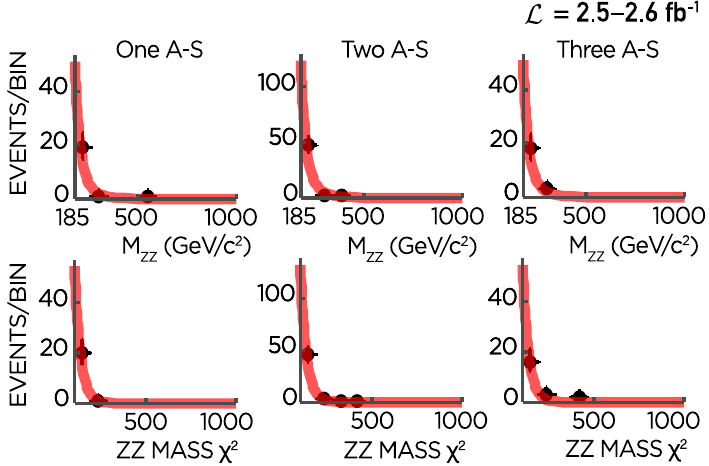}
  } \caption{$M_{\mu\mu\mu\mu}$ and $\chi^2_{ZZ}$ for the one-, two-,
  and three-anti-selected (A-S) four-muon samples, and the results of the
  simultaneous non-resonant background shape fit to the two- and
  three-anti-selected muon samples.}\label{fig:mmmm_fit_projections}
\end{figure}

The background shapes obtained from these fits are normalized so that
the sums of the integrals for $185 < M_{\Xboson} < 300\,\gevcc$ and
the simulation-derived diboson predictions match the number of events observed
with $185 < M_{\Xboson} < 300\,\gevcc$ in the four-lepton samples. The
shapes are then extrapolated into the low $\chi^2_{ZZ}$, high
$M_{\Xboson}$ $\dil\dil$ signal regions. The statistical uncertainty
on the normalization is the dominant source of uncertainty for the
four-lepton non-resonant background prediction.

As one test of the independence of the non-resonant predictions to the
number of selected/anti-selected leptons, Tables
\ref{table:llll_shape_check} and \ref{table:llll_shape_check_yield} show the parameters and yield
predictions obtained by fitting $\gamma$ for each sample independently
of the others. All of the yield predictions for a given signal mass and decay channel are consistent with each
other within the statistical uncertainty.

\begin{table}
  \centering \caption{Comparison of non-resonant background mass shape
  parameter $\gamma$ fitted independently in the individual sideband
  samples with the simultaneous fit to the two- and three-anti-selected lepton samples.}
  \begin{tabular}{cD..{3.2}cD..{1.2}D..{3.2}cD..{1.2}}
     \multicolumn{1}{c}{ Channel} &
     \multicolumn{3}{c}{\EEEE} & 
     \multicolumn{3}{c}{\EEMM} \\ \hline \hline
    \noalign{\vspace{1mm}} 
    Simultaneous & -4.39 & \PM & 0.09 & -5.42 & \PM & 0.15 \\ 
    3 anti-leptons &  -4.22 &\PM& 0.11 & -5.19 &\PM& 0.17  \\
    2 anti-leptons & -4.50 &\PM& 0.15 & -4.70 &\PM& 0.21  \\
    1 anti-lepton & -4.21 &\PM& 0.30 & -3.63 &\PM& 0.32  \\
    0 anti-leptons & -5.04 &\PM& 0.94 & -3.54 &\PM& 0.78  \\ \hline \hline
  \end{tabular} \\  
  \begin{tabular}{cD..{3.2}cD..{1.2}D..{3.2}cD..{1.2}}
     \multicolumn{1}{c}{ Channel} &
     \multicolumn{3}{c}{\MMEE} & 
     \multicolumn{3}{c}{\MMMM} \\ \hline \hline
    \noalign{\vspace{1mm}} 
    Simultaneous & -5.25  &\PM& 0.34 & -6.51 &\PM& 0.61 \\ 
    3 anti-leptons & -4.98 &\PM& 0.50  & -6.3  &\PM& 1.1 \\
    2 anti-leptons & -4.96 &\PM& 0.41 & -6.60  &\PM& 0.73 \\
    1 anti-lepton & -5.25 &\PM& 0.73 & -5.33  &\PM& 0.92 \\
    0 anti-leptons & -5.7  &\PM& 1.6  & -4.4  &\PM& 1.3 \\  \hline \hline
  \end{tabular}
  \label{table:llll_shape_check}
\end{table}

\begin{table}
  \centering \caption{Non-resonant background
  predictions for a characteristic example (the 410--590 \gevcc
  four-body mass range appropriate for a 500 \gevcc signal) from fits
  of the individual sideband samples, compared with the prediction from the
  simultaneous fit to the two- and three-anti-selected lepton
  samples.}
  \begin{tabular}{cD..{3.2}cD..{1.2}D..{3.2}cD..{1.2}}
     \multicolumn{1}{c}{ Channel} &
     \multicolumn{3}{c}{\EEEE} & 
     \multicolumn{3}{c}{\EEMM} \\ \hline \hline
    \noalign{\vspace{1mm}} 
    Simultaneous   & 0.64 &\PM& 0.29 & 0.128 &\PM& 0.064  \\ 
    3 anti-leptons & 0.65 &\PM& 0.31 & 0.093 &\PM& 0.048  \\
    2 anti-leptons & 0.54 &\PM& 0.26 & 0.166 &\PM& 0.091  \\
    1 anti-lepton  & 0.58 &\PM& 0.30 & 0.40  &\PM& 0.24   \\
    0 anti-leptons & 0.51 &\PM& 0.35 & 0.82  &\PM& 0.59   \\ \hline \hline
  \end{tabular} \\  
  \begin{tabular}{cD..{3.2}cD..{1.2}D..{3.2}cD..{1.2}}
     \multicolumn{1}{c}{ Channel} &
     \multicolumn{3}{c}{\MMEE} & 
     \multicolumn{3}{c}{\MMMM} \\ \hline \hline
    \noalign{\vspace{1mm}} 
    Simultaneous   & 0.130  &\PM& 0.077 & 0.063 &\PM& 0.040 \\ 
    3 anti-leptons & 0.094  &\PM& 0.060 & 0.058 &\PM& 0.044 \\
    2 anti-leptons & 0.127  &\PM& 0.078 & 0.064 &\PM& 0.043 \\
    1 anti-lepton  & 0.095  &\PM& 0.065 & 0.17  &\PM& 0.13 \\
    0 anti-leptons & 0.14   &\PM& 0.12  & 4.2   &\PM& 3.8 \\  \hline \hline
  \end{tabular}
  \label{table:llll_shape_check_yield}
\end{table}

The sideband data fit for the $\dil jj$ non-resonant background
estimates consist of events containing a dilepton pair with
$\chi^2_{Z} < 25$ and a dijet pair with either $40 < M_{jj} <
65\,\gevcc$ or $120 < M_{jj} < 200\,\gevcc$. The $M_{jj}$ spectrum
near the \Zboson boson pole mass is exponentially falling before imposing
any requirement on $M_{\Xboson}$ but linear for events with
$M_{\Xboson} > 300\,\gevcc$, where the effect of the four-body mass
cut is to sculpt a peak in the dijet mass at $M_{jj} \gg 200 \,\gevcc$
(see Fig. \ref{fig:eeqq_lowmx}). We linearly interpolate the
background expectation for $65 < M_{jj} < 120\,\gevcc$ from the lower
and higher $M_{jj}$ sideband data. To avoid underestimating the
background at very high $M_{\Xboson}$ where these sidebands are empty,
we model the population of either sideband vs $M_{\Xboson}$ with an
exponential fit to the available data to obtain the numbers used in
the interpolation. Fig. \ref{fig:ll_dijet_mass_extrap} shows the numbers of events in the two dijet
mass sidebands as a function of the requirement on minimum four-body
mass and the exponential fits. Exponential functions model the data well.

\begin{figure} \centering
  \mbox{ \includegraphics{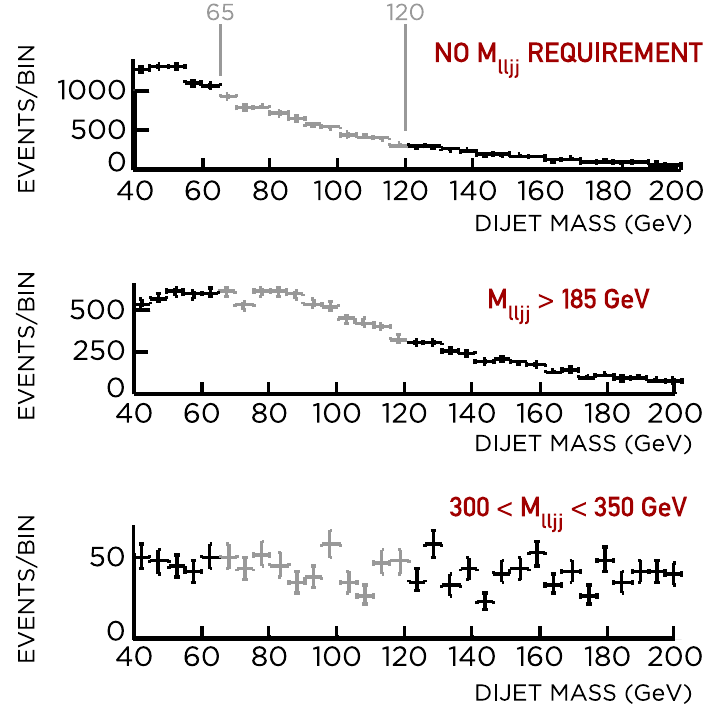}
  } \caption{Dijet mass spectra for two-electron two-jet candidates
  with $\chi^2_{ee} < 25$ and three $M_{eejj}$ requirements: no requirement, 
  $M_{eejj} > 185\,\gevcc$, and $300 < M_{eejj} < 350\,\gevcc$,
  beyond which the shape of the dijet mass in the \Zboson boson region is
  linear.}\label{fig:eeqq_lowmx}
\end{figure}

\begin{figure} \centering
  \mbox{ \includegraphics[width=7.2cm]{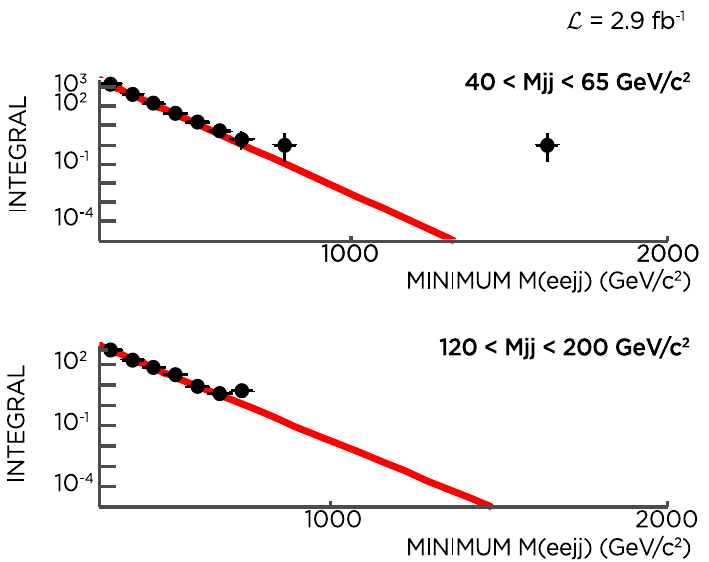}}
  \mbox{ \includegraphics[width=7.2cm]{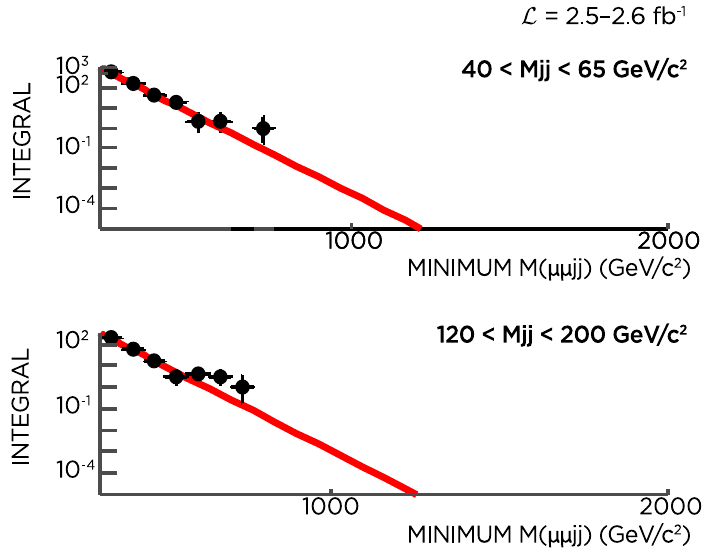}}
  \caption{Number of two-lepton two-jet events with $40 < M_{jj} <
  65\,\gevcc$ and $120 < M_{jj} < 200\,\gevcc$ as a function of the
  minimum $M_{lljj}$ mass requirement, along with the exponential fit
  to each that is used to interpolate the background prediction for
  the $65 < M_{jj} < 120\,\gevcc$
  region.}\label{fig:ll_dijet_mass_extrap}
\end{figure}

As one unbiased test of the prediction of the dijet mass spectrum, we
repeat the selection and fit procedure on samples consisting of events
containing a trigger lepton plus an anti-selected lepton and at least
two jets. Comparison of the fit predictions against these $65 <
M_{jj} < 120\,\gevcc$ data, which are depleted of signal, show the
method performs well (Tables \ref{table:efqq_checks} and
\ref{table:mfqq_checks}). The disagreement in the lowest $M_{eejj}$
bin for this and other control samples is a result of a slight
deviation from an exponential distribution. This residual variation is taken as an
extra systematic uncertainty in the lowest mass bin.

\begin{table}
  \caption{Comparison of dijet mass fit predictions with data for the
  trigger electron plus anti-selected electron and two-jet channel
  sample. Shown are the uncertainties on the mean
  prediction due to fit statistics.}\label{table:efqq_checks}
  \begin{center} \begin{tabular}{cD..{5.2}cD..{2.2}r}
  \multicolumn{1}{c}{$M_{ee jj}$ ({$\gevcc$})} &
   \multicolumn{3}{c}{Prediction} & 
   {Observed}\\ \hline \hline
   \noalign{\vspace{1mm}} 
   350--450 &  1105 & \PM & 29  & 941 \\ 
   400--600 &  488 & \PM & 21  & 540 \\ 
   500--700 &  95.6 & \PM & 7.2 & 115 \\
    600--800 &  18.8 & \PM & 2.0  & 18 \\
  650--950 &  8.31 & \PM & 1.1  & 8 \\ 
  750--1050 & 1.64& \PM & 0.26  & 3 \\
   800--1200 & 0.73& \PM & 0.13  & 2 \\ \hline \hline
   \end{tabular} \end{center}
\end{table}

\begin{table}
  \caption{Comparison of dijet mass fit predictions with data for the
  trigger muon plus anti-selected muon and two-jet channel
  sample. Shown are the uncertainties on the mean
  prediction due to fit statistics.}\label{table:mfqq_checks} 
  \begin{center} \begin{tabular}{cD..{3.4}cD..{2.4}r}
  \multicolumn{1}{c}{$M_{\mu\mu jj}$ ({$\gevcc$})} &
   \multicolumn{3}{c}{Prediction} & 
   {Observed}\\ \hline \hline
   \noalign{\vspace{1mm}} 
   350--450 & 23.7 & \PM & 4.0 & 22 \\ 
   400--600 &  9.4 & \PM & 2.2  & 9 \\ 
   500--700 &  1.48 & \PM & 0.59 & 2 \\ 
   600--800 &  0.23 & \PM & 0.15  & 1 \\ 
   650--950 &  0.093 & \PM & 0.072  & 1 \\ 
   750--1050 & 0.015& \PM & 0.017  & 0 \\
   800--1200 & 0.0059& \PM & 0.0084  & 0 \\  \hline \hline
   \end{tabular} \end{center}
\end{table}

We determine the backgrounds resonant in both \Zboson boson masses with
Monte Carlo models normalized to the cross-sections predicted by
{\sc mcfm}~\cite{Campbell:1999p576}. In the $\dil jj$ channels, our dijet
mass resolution and analysis selection does not distinguish between
the \WZQL and \ZZJL, and so the background from both processes is
present.

\begin{table}
  \caption{Total $eeee$ backgrounds with $\ZZMassChisq < 50$ for each
  signal mass \MX. The uncertainty includes the \ZZ diboson production cross-section
  uncertainty, the luminosity uncertainty, the statistical uncertainty
  from the simulation, and the statistical uncertainty from the
  non-resonant background fit.}
  \label{table:resonant_bkgs_eeee} 
  \begin{center} 
  \begin{tabular}{rD..{3.4}cD..{1.4}D..{3.2}cD..{1.2}}
  \multicolumn{1}{c}{{\MX}} &
  \multicolumn{3}{c}{\multirow{2}{*}{ SM \ZZ}} & 
  \multicolumn{3}{c}{\multirow{2}{*}{ Non-resonant}} \\ 
  \multicolumn{1}{c}{{ ($\gevcc$)}} & & & & & & \\ \hline \hline
  \noalign{\vspace{1mm}} 
  400 & 0.22 & \PM & 0.02 & 1.31 & \PM & 0.44 \\ 
  500 & 0.086 & \PM & 0.009 & 0.64 & \PM & 0.29 \\ 
  600 & 0.045 & \PM & 0.005 & 0.44 & \PM & 0.18 \\ 
  700 & 0.020 & \PM & 0.003 & 0.28 & \PM & 0.20 \\ 
  800 & 0.007 & \PM & 0.001 & 0.15 & \PM & 0.12 \\ 
  900 & 0.005 & \PM & 0.001 & 0.14 & \PM & 0.11 \\ 
  1000 & 0.0032 & \PM & 0.0001 & 0.11 & \PM & 0.10 \\ \hline \hline
  \end{tabular} \end{center}
\end{table}

\begin{table}
  \caption{Total $ee\mu\mu$ backgrounds with $\ZZMassChisq < 50$ for each
  signal mass \MX. The uncertainty includes the \ZZ diboson production cross-section
  uncertainty, the luminosity uncertainty, the statistical uncertainty
  from the simulation, and the statistical uncertainty from the
  non-resonant background fit.}
  \label{table:resonant_bkgs_eemm}
  \begin{center} 
  \begin{tabular}{rD..{3.4}cD..{1.4}D..{3.3}cD..{1.3}}
  \multicolumn{1}{c}{{\MX}} &
  \multicolumn{3}{c}{\multirow{2}{*}{ SM \ZZ}} & 
  \multicolumn{3}{c}{\multirow{2}{*}{ Non-resonant}} \\ 
  \multicolumn{1}{c}{{ ($\gevcc$)}} & & & & & & \\ \hline \hline
  \noalign{\vspace{1mm}} 
    400  &      0.19 & \PM & 0.02  &     0.33 & \PM & 0.13 \\
    500  &      0.067 & \PM & 0.007 & 0.128 & \PM & 0.064 \\    
    600  &      0.035 & \PM & 0.004 &     0.075 & \PM & 0.047 \\   
    700  &      0.014 & \PM & 0.002 &     0.041 & \PM & 0.029 \\   
    800  &      0.004 & \PM & 0.001 &    0.019 & \PM & 0.013 \\   
    900  &      0.003 & \PM & 0.001 &     0.016 & \PM & 0.013 \\   
    1000 &        0.0013 & \PM & 0.0006 &    0.012 & \PM & 0.011 \\   \hline \hline
  \end{tabular} \end{center}
\end{table}

\begin{table}
  \caption{Total $\mu\mu ee$ backgrounds with $\ZZMassChisq < 50$ for each
  signal mass \MX. The uncertainty includes the \ZZ diboson production cross-section
  uncertainty, the luminosity uncertainty, the statistical uncertainty
  from the simulation, and the statistical uncertainty from the
  non-resonant background fit.}
  \label{table:resonant_bkgs_mmee} 
  \begin{center}
  \begin{tabular}{rD..{3.4}cD..{1.4}D..{3.3}cD..{1.3}}
  \multicolumn{1}{c}{{\MX}} &
  \multicolumn{3}{c}{\multirow{2}{*}{ SM \ZZ}} & 
  \multicolumn{3}{c}{\multirow{2}{*}{ Non-resonant}} \\ 
  \multicolumn{1}{c}{{ ($\gevcc$)}} & & & & & & \\ \hline \hline 
  \noalign{\vspace{1mm}} 
    400  &      0.077 & \PM & 0.008  & 0.32 & \PM & 0.16 \\
    500  &      0.027 & \PM & 0.003  & 0.130 & \PM & 0.077 \\   
    600  &      0.014 & \PM & 0.002  & 0.078 & \PM & 0.055 \\  
    700  &      0.0065 & \PM & 0.0010 & 0.044 & \PM & 0.034 \\  
    800  &      0.0018 & \PM & 0.0007 & 0.021 & \PM & 0.017 \\  
    900  &      0.0014 & \PM & 0.0006 & 0.018 & \PM & 0.017 \\
    1000 &      0.0011 & \PM & 0.0005 & 0.014 & \PM & 0.013 \\  \hline \hline
  \end{tabular} \end{center}
\end{table}

\begin{table}
  \caption{Total $\mu\mu\mu\mu$ backgrounds with $\ZZMassChisq < 50$ for each
  signal mass \MX. The uncertainty includes the \ZZ diboson production cross-section
  uncertainty, the luminosity uncertainty, the statistical uncertainty
  from the simulation, and the statistical uncertainty from the
  non-resonant background fit.}
  \label{table:resonant_bkgs_mmmm}
  \begin{center}
  \begin{tabular}{rD..{2.5}cD..{1.5}D..{1.4}cD..{1.4}}
  \multicolumn{1}{c}{{\MX}} &
  \multicolumn{3}{c}{\multirow{2}{*}{SM \ZZ}} & 
  \multicolumn{3}{c}{\multirow{2}{*}{Non-resonant}} \\ 
  \multicolumn{1}{c}{{($\gevcc$)}} & & & & & & \\ \hline \hline 
  \noalign{\vspace{1mm}} 
    400   &     0.090 & \PM & 0.010  &       0.21 & \PM & 0.11 \\
    500   &     0.036 & \PM & 0.005  &       0.063& \PM & 0.040 \\     
    600  &      0.018 & \PM & 0.002 &        0.031& \PM & 0.023 \\     
    700    &    0.0082 & \PM & 0.0015  &     0.015& \PM &0.013 \\     
    800    &    0.0018 & \PM & 0.0007  &      0.0056& \PM & 0.0049 \\    
    900     &   0.00011 & \PM & 0.00005 &      0.0046& \PM &0.0042  \\   
    1000 &        0.0009 & \PM & 0.0005 &       0.0031& \PM &0.0030     \\  \hline \hline
  \end{tabular} \end{center}
\end{table}

\begin{table*}
  \caption{Total $ee jj$ backgrounds with $65 < M_{jj} <
  120\,\gevcc$ for each signal mass \MX. The uncertainty includes diboson cross-section
  uncertainties, the uncertainty on the luminosity, the statistical uncertainty from the
  simulation, and the uncertainties from the non-resonant background fits.}
  \label{table:resonant_bkgs_eejj}
  \begin{center} 
  \begin{tabular}{rD..{3.2}cD..{1.2}D..{3.3}cD..{1.3}D..{5.2}cD..{2.2}}
  \multicolumn{1}{c}{{\MX}} &
  \multicolumn{3}{c}{\multirow{2}{*}{ SM \ZZ}} & 
  \multicolumn{3}{c}{\multirow{2}{*}{ SM \WZ}} & 
  \multicolumn{3}{c}{\multirow{2}{*}{ Non-resonant}} \\ 
  \multicolumn{1}{c}{{ ($\gevcc$)}} & & & & & & \\ \hline \hline
  \noalign{\vspace{1mm}} 
    400  &     5.72 & \PM & 0.97 & 9.4&  \PM & 1.1  & 483  & \PM &18  \\
    500  &     2.43 & \PM & 0.58 & 3.25& \PM & 0.46 & 128.0& \PM &8.2  \\   
    600  &     0.99 & \PM & 0.36 & 1.10& \PM & 0.22 & 47.4 & \PM &4.1  \\  
    700  &     0.19 & \PM & 0.18 & 0.60& \PM & 0.16 & 14.9 & \PM &1.7 \\   
    800  & \multicolumn{3}{c}{$0^{+0.11}$} &  0.158 & \PM & 0.083 & 2.86 & \PM &0.46   \\ 
    900  & \multicolumn{3}{c}{$0^{+0.11}$} &  0.095 & \PM & 0.067 & 1.75 & \PM &0.31  \\  
    1000 & \multicolumn{3}{c}{$0^{+0.11}$} &  \multicolumn{3}{c}{$0^{+0.067}$} & 0.77& \PM &0.16 \\     \hline \hline
  \end{tabular} \end{center}
\end{table*}

\begin{table*}
  \caption{Total $\mu\mu jj$ backgrounds with $65 < M_{jj} <
  120\,\gevcc$ for each signal mass \MX. The uncertainty includes diboson cross-section
  uncertainties, the uncertainty on the luminosity, the statistical uncertainty from the
  simulation, and the uncertainties from the non-resonant background fits.}
  \label{table:resonant_bkgs_mmjj}
  \begin{center} 
  \begin{tabular}{rD..{3.2}cD..{1.2}D..{3.3}cD..{1.3}D..{5.3}cD..{2.3}}
  \multicolumn{1}{c}{{\MX}} &
  \multicolumn{3}{c}{\multirow{2}{*}{ SM \ZZ}} & 
  \multicolumn{3}{c}{\multirow{2}{*}{ SM \WZ}} & 
  \multicolumn{3}{c}{\multirow{2}{*}{ Non-resonant}} \\ 
  \multicolumn{1}{c}{{ ($\gevcc$)}} & & & & & & \\ \hline \hline
  \noalign{\vspace{1mm}} 
   400   &    2.90 & \PM & 0.57 &      6.04 & \PM & 0.73 &      162 & \PM & 11  \\  
   500   &    1.30 & \PM & 0.38  &     2.06 & \PM & 0.32 &     37.7 & \PM & 4.4  \\  
   600   &    0.57 & \PM & 0.26  &     0.73 & \PM & 0.17  &     12.6 & \PM & 2.0  \\  
   700   &    0.26 & \PM & 0.19  &     0.229 & \PM & 0.93 &     3.53 & \PM & 0.72  \\
   800   &    0.09 & \PM & 0.13  &     0.023 & \PM & 0.040 &    0.57 & \PM & 0.16     \\
   900   &\multicolumn{3}{c}{$0^{+0.10}$} &\multicolumn{3}{c}{$0^{+0.032}$} & 0.33 & \PM & 0.10 \\
   1000  &\multicolumn{3}{c}{$0^{+0.10}$} &\multicolumn{3}{c}{$0^{+0.032}$} & 0.133 & \PM & 0.045 \\ \hline \hline
  \end{tabular} \end{center}
\end{table*}

Tables \ref{table:resonant_bkgs_eeee}
through \ref{table:resonant_bkgs_mmjj} show the total prediction for
each analysis channel. At each signal mass, the predictions are
integrated over the four-body mass range listed in
Table \ref{table:signal_bins}. The uncertainties listed for the
diboson predictions consist of the error on the {\sc mcfm}
cross-section ($\approx 7\%$), the uncertainty on the luminosity
(6\%), and the statistical uncertainty due to finite Monte Carlo
statistics, which is the dominant uncertainty on the diboson
prediction at high four-body mass, though a negligble component of the
total background uncertainty. The uncertainties listed for the
non-resonant backgrounds consist of the statistical uncertainty from
the shape parameters and the normalization uncertainty due to the
small number of events in the $185 < M_{\Xboson} < 300\,\gevcc$
four-lepton control regions. The non-resonant and diboson background
systematic uncertainties are negligible compared to the statistical uncertainty on
the total background.

\section{Results}\label{sec:results}

We optimized all selections and estimated all backgrounds before
examining the data with $M_{llll} > 300\,\gevcc$ or with a dilepton pair
having $\chi^2_{Z} < 25$ and a dijet pair with $65 < M_{jj} <
120\,\gevcc$. Figs. \ref{fig:llll_summary} and \ref{fig:lljj_summary}
show the data in these regions and the combined resonant and
non-resonant background predictions for all four-lepton channels and
for both dijet channels. In all cases the data agree with the total
background prediction and provide no compelling evidence for
resonant \ZZ diboson production. The highest-mass $\dil\dil$ event
(577 \gevcc) consists of four muons. For this event, one \Zboson boson 
candidate has a mass of $79 \pm 4.2\,\gevcc$ and the other \Zboson boson 
candidate has a mass of $400 \pm 170\,\gevcc$. The large mass
uncertainty of the second \Zboson boson comes from a large curvature
uncertainty in the measurement of one $\pT = 290\,\gevc$ muon with few
COT hits. The highest-mass $\dil jj$ event (868 \gevcc) has $M_{ee} =
96.5 \pm 1.3\,\gevcc$ and $M_{jj} = 77.8 \pm 6.5\,\gevcc$.

\begin{figure} \centering
  \mbox{ \includegraphics[width=7.2cm]{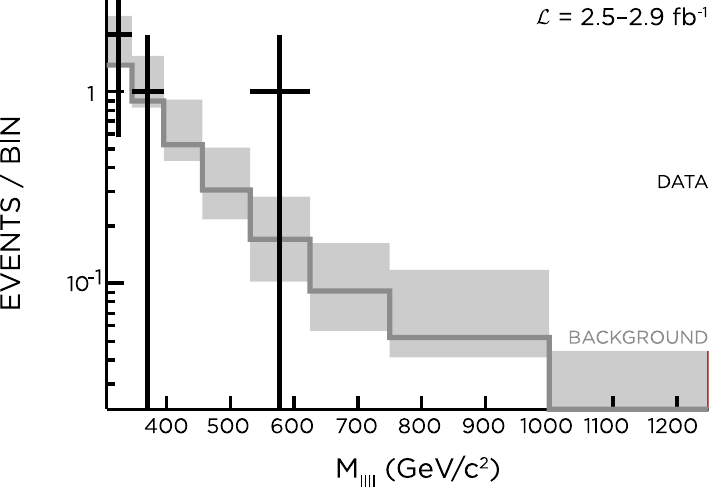}
  } \caption{Prediction and data for all four-lepton
  channels combined. The background prediction for each bin consists
  of the integral of the non-resonant background functions and diboson
  Monte Carlo determined in Section \ref{sec:background}. The
  background predictions agree with the data.}\label{fig:llll_summary}
\end{figure}

\begin{figure} \centering
  \mbox{ \includegraphics[width=7.2cm]{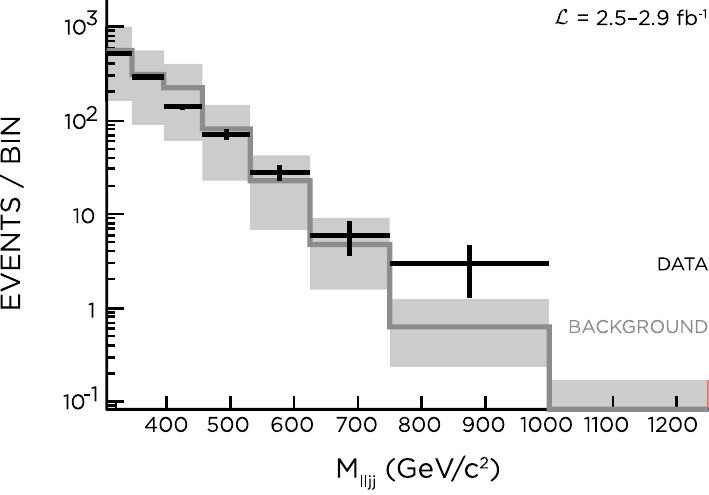}
  } \caption{Prediction and data for both $\dil jj$ channels
  combined. The background prediction for each bin consists of the
  integral of the non-resonant background functions and diboson Monte
  Carlo determined in Section \ref{sec:background}. The background
  predictions agree with the data. }\label{fig:lljj_summary}
\end{figure}

Absent evidence of a signal, we communicate our sensitivity to $\XZZ$
processes by setting limits with an acceptance from a
widely-available \herwig Monte Carlo process \cite{Corcella:2000bw},
the spin-2 Kaluza-Klein graviton. The total acceptance times
efficiency for this process varies between roughly 40-50\% for a given
four-lepton channel and between 20-40\% for a given dijet channel
(Fig. \ref{fig:acceff_vs_mass}). At higher masses, the fraction of \Zboson boson
decays increases in which the angular separation between products is
too small for the calorimeter to resolve. This lowers the acceptances
for the dijet modes and, to a lesser extent, the electron modes.

The combined effect of the lepton reconstruction and identification
improvements on graviton signal is demonstrated in
Figs. \ref{fig:eeee_yield_comparison} and
\ref{fig:mmmm_yield_comparison} for the four-electron and four-muon
channels. We compute data yields and estimates of signal and
background by integrating the Monte Carlo predictions and fitted
non-resonant background shapes over a set of overlapping,
variable-width bins for signal masses from 400 \gevcc to 1 \tevcc (Table
\ref{table:signal_bins}). Each signal bin width is chosen to be large
enough to fully contain the four-body mass distribution expected for
an intrinsically narrow signal and the broadening from systematic
effects. Table \ref{table:limit_yields} shows the total background
prediction and observed data yields in each of these bins.

\begin{figure} \centering
  \mbox{ \includegraphics{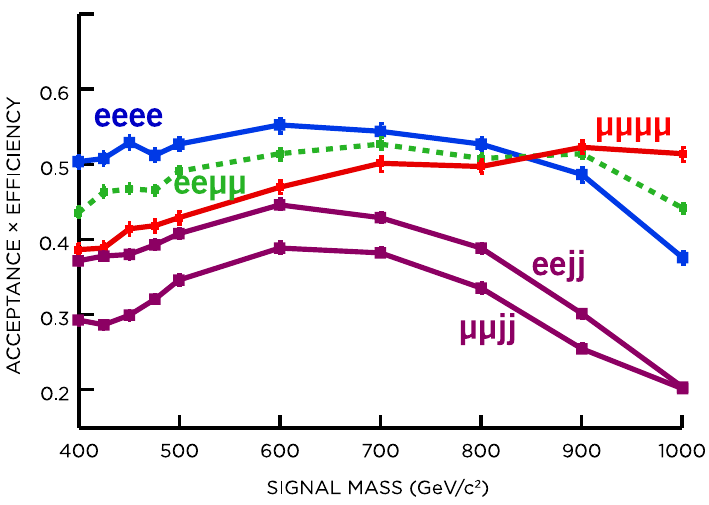}} \caption{Products
  of acceptance times efficiency for each of the graviton analysis
  channels and their dependence on graviton mass. These do not include
  $\Zboson \Zboson$ diboson branching ratios, which for each
  $\Zboson \Zboson \rightarrow \dil jj$ mode are approximately 40
  times the branching ratios for $\Zboson \Zboson \rightarrow eeee$ or
  $\Zboson \Zboson \rightarrow \mu\mu\mu\mu$. The $ee \mu\mu$ and
  $\mu\mu ee$ acceptances have been summed. }
  \label{fig:acceff_vs_mass}
\end{figure}

\begin{figure} \centering
  \mbox{ \includegraphics[width=7.2cm]{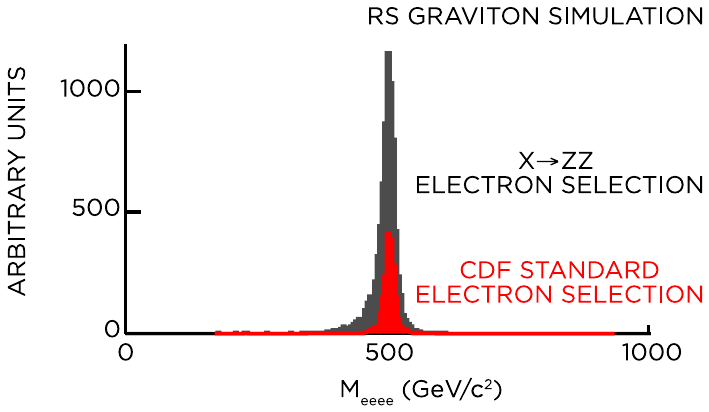} }
  \caption{Four-electron yield comparison for a 500 $\gevcc$
    graviton between (lower histogram) the CDF standard electron selection
    criteria and (upper histogram) the criteria employed in the present
    analysis.}\label{fig:eeee_yield_comparison}
\end{figure}

\begin{figure} \centering
  \mbox{ \includegraphics[width=7.2cm]{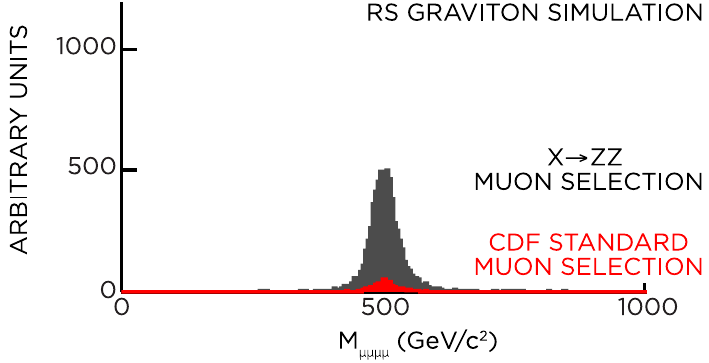} }
  \caption{Four-muon yield comparison for a 500 $\gevcc$
    graviton between (lower histogram) a CDF standard muon selection
    criteria and (upper histogram) the criteria employed in the present
    analysis.}\label{fig:mmmm_yield_comparison}
\end{figure}

\begin{table}
  \caption{Signal binning used for limit-setting.}
  \label{table:signal_bins}
  \begin{center}
    \begin{tabular}{rcr}
      \multicolumn{1}{c}{Signal Mass} & & 
      \multicolumn{1}{c}{Bin Half} \\
      \multicolumn{1}{c}{($\gevcc$)} & & \multicolumn{1}{c}{ Width ($\gevcc$)}\\ \hline \hline
   \noalign{\vspace{1mm}} 
      400 & \PM & 70 \\
      500 & \PM & 90 \\
      600 & \PM & 130 \\
      700 & \PM & 160 \\
      800 & \PM & 160 \\
      900 & \PM & 230 \\
      1000 & \PM & 280 \\\hline \hline
    \end{tabular}
  \end{center}
\end{table}

\begin{table*}
   \caption{Total background prediction and observed data yields for
   each of the limit-setting bins in
   Table \ref{table:signal_bins}. Successive bins are partially
   correlated. The uncertainty (quoted as the least two significant
   figures in parentheses) is the systematic uncertainty on the mean
   background prediction and does not include statistical fluctuation
   about the mean.}
   \label{table:limit_yields}
   \begin{center}
     \begin{tabular}{ccccccccc}
       \multirow{2}{*}{Channel} & & \multicolumn{7}{c}{{\MX (\gevcc)}} \\
       & & 400 & 500 & 600 & 700 & 800 & 900 & 1000 \\ \hline \hline
       \multirow{2}{*}{$ee ee$} & Expected & 1.53(44) & 0.73(29) & 0.49(18) & 0.30(20) & 0.16(12) & 0.15(11) & 0.11(10) \\
                                & Observed & 0        & 0        & 0        & 0        & 0        & 0        & 0 \\ \hline
       \multirow{2}{*}{$ee \mu\mu$} & Expected & 0.52(13) & 0.195(64) & 0.110(47) & 0.055(29) & 0.023(13) & 0.019(13) & 0.013(11) \\
                                    & Observed & 2        & 0         & 0         & 0         & 0         & 0         & 0 \\ \hline
       \multirow{2}{*}{$\mu\mu ee$} & Expected & 0.397(16) & 0.157(77) & 0.092(55) & 0.050(34) & 0.023(17) & 0.019(17) & 0.015(13) \\
                                    & Observed & 0        & 0         & 0         & 0         & 0         & 0         & 0 \\ \hline
       \multirow{2}{*}{$\mu\mu \mu\mu$} & Expected & 0.30(11) & 0.099(40) & 0.049(23) & 0.023(13) & 0.0074(49) & 0.0047(42) & 0.0040(30) \\
                                        & Observed & 1        & 1         & 1         & 1         & 0         & 0         & 0 \\ \hline
       \multirow{2}{*}{$ee jj$} & Expected & 498(18) & 133.7(82) & 49.5(41) & 15.7(17) & 3.02(48) & 1.84(34) & 0.77(21) \\
                                & Observed & 456     & 142       & 69       & 28       & 7        & 5        & 2      \\ \hline
       \multirow{2}{*}{$\mu\mu jj$} & Expected & 171(11) & 41.1(44)  & 13.9(20) & 4.02(75) & 0.68(21) & 0.33(14) & 0.13(11) \\
                                    & Observed & 143     & 41        & 19       & 4        & 2        & 2        & 1 \\ \hline \hline
     \end{tabular}
   \end{center}
\end{table*}

We calculate 95\%-credibility upper limits as a function of signal
mass using a six-channel product of Bayesian likelihoods and a uniform
prior for the (non-negative) $\Xboson \rightarrow ZZ$
cross-section. We use marginalized truncated-Gaussian nuisance
parameters for the luminosity, background predictions, and signal
efficiencies, and we account for systematic uncertainties correlated
amongst the six channels when appropriate. As discussed earlier, we
assign a 20\% uncorrelated uncertainty to the total acceptance
$\times$ efficiency for each channel to account for the observed
time-dependent variation in the Drell-Yan $\mu\mu$ cross-section,
conservatively covering the sum of individual systematic uncertainties
such as signal acceptance uncertainties in order to simplify the
combination. Studies of the individual uncertainties indicate the
largest contribution after the uncertainty due to the \ZMM cross
section variation is the 5.9\% uncertainty on the luminosity. In
addition to the observed limit, we compute expected limits from 10,000
pseudo-experiments at each candidate \Xboson
mass. Fig. \ref{fig:limit} shows the resultant limits along with the
$k/M_{Pl} = 0.1$ Randall-Sundrum (RS1) graviton cross-section from {\sc
herwig}. The present search improves the ${\cal{O}}(\text{4 \pb})$
limit of the earlier $eeee$ search \cite{CDFCollaboration:2008p1237}
by an order of magnitude.

\begin{figure} \centering
  \mbox{ \includegraphics{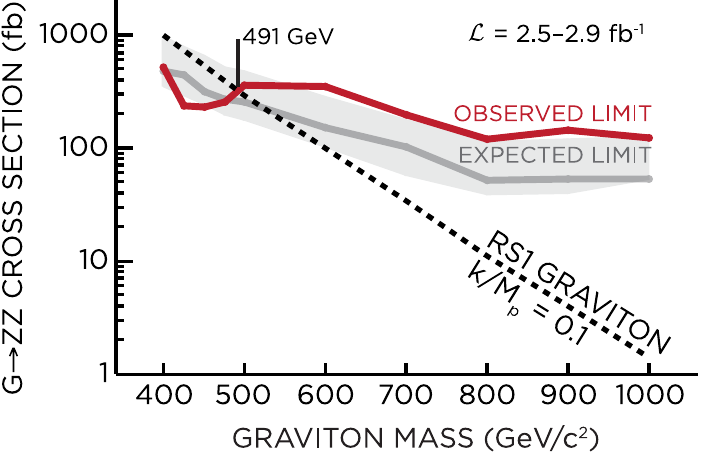}
  } \caption{95\%-credibility cross-section upper limit assuming acceptance
  for a massive graviton along with the limit and 68\% variation
  expected for the background-only hypothesis.}\label{fig:limit}
\end{figure}

\section{Conclusions}\label{sec:conclusions}
We have reported on an improved search for a massive resonance decaying
to \ZZ dibosons via the $eeee$, $ee\mu\mu$, $\mu\mu\mu\mu$, $eejj$, and $\mu\mu
jj$ channels. We find that the four-body invariant mass spectrum above
$300\,\gevcc$ is consistent with background estimates derived from
sideband data samples and electroweak Monte Carlo models. To quantify
our sensitivity, we set limits using the acceptance for a
Randall-Sundrum graviton model that are 7--20 times stronger than the
previously published direct limits on resonant \ZZ diboson production.

\begin{acknowledgments}
We thank the Fermilab staff and the technical staffs of the
participating institutions for their vital contributions. This work
was supported by the U.S. Department of Energy and National Science
Foundation; the Italian Istituto Nazionale di Fisica Nucleare; the
Ministry of Education, Culture, Sports, Science and Technology of
Japan; the Natural Sciences and Engineering Research Council of
Canada; the National Science Council of the Republic of China; the
Swiss National Science Foundation; the A.P. Sloan Foundation; the
Bundesministerium f\"ur Bildung und Forschung, Germany; the Korean
World Class University Program, the National Research Foundation of
Korea; the Science and Technology Facilities Council and the Royal
Society, UK; the Institut National de Physique Nucleaire et Physique
des Particules/CNRS; the Russian Foundation for Basic Research; the
Ministerio de Ciencia e Innovaci\'{o}n, and Programa
Consolider-Ingenio 2010, Spain; the Slovak R\&D Agency; and the
Academy of Finland.
\end{acknowledgments}

\appendix*
\begin{prdappendix}\label{appendix:tracking}

The standard CDF reconstruction software uses two main approaches to
reconstruct tracks. High quality central tracking ($|\eta|<1$) starts
in the COT and assembles piecewise segments of up to 12 hits in each
superlayer, fits them, and groups them into tracks to which any
available silicon hits are then attached in an outside-in
search. Afterward, ``silicon standalone'' tracking starts with all
possible combinations of three unused silicon hits, searches the
remaining silicon layers, and projects successful tracks into the COT
to attach any compatible hits in order to improve the track momentum
resolution and lower the fake rate.

The combination of these approaches results in low efficiency in the
$1 < |\eta| < 2$ region. Tracks originating from $z=0$ with $|\eta| <
1.7$ will leave traces of their passage in the lowest-radii
superlayers of the COT. Though very efficient when full COT coverage
is available, for $|\eta|>1$ the central tracking algorithms lose
efficiency nearly linearly with $|\eta|$ reaching zero efficiency at about 
$|\eta|=1.6$. The silicon fully covers $|\eta|<1.8$ to compensate for
the falling COT efficiency, but the existing silicon-driven tracking
algorithms reconstruct tracks with low efficiency and produce
low-quality or spurious tracks with poor pointing resolution into the
COT. Thus the COT information for forward tracks is rarely exploited.

This analysis employs a thorough revision of the forward and central
tracking algorithms in order to reconstruct tracks with better
efficiency and resolution, including a new ``Backward'' algorithm that
makes full use of the partial COT coverage. The Backward algorithm, illustrated in Fig.
\ref{fig:bw-example} for a simple case, starts
by searching the COT for hits unused by the central COT algorithm and
constructing segments in one of the inner axial superlayers consisting
of no more than 12 hits. At this stage, the position measurements
contain a drift sign ambiguity and important drift time corrections,
such as large time of flight and sense wire signal propagation times,
are unknown and cannot be approximated by the constant corrections
assumed for the central segment pattern recognition. The Backward
algorithm solves this problem with a variant of the central segment
pattern recognition that resolves the drift sign ambiguity and drift
time corrections during the search and is optimized for tracking in
the low radius, high hit density inner superlayers and near the COT
endplates. Once unused COT hit segments are found consistent with a
forward track, the algorithm then fits the segments with a beamline
constraint to obtain five-parameter helices that intersect the $z$
position of the highest sum $\pT$ $z$ vertex identified using central
algorithm tracks. In most cases, the fits do not conclusively identify
stereo COT measurements in the innermost stereo superlayer, and so
multiple helices are obtained corresponding to trajectories through
each possible combination of silicon module and polar angle. Drift
time corrections are recomputed for each case.

\begin{figure} \centering
  \mbox{ \includegraphics{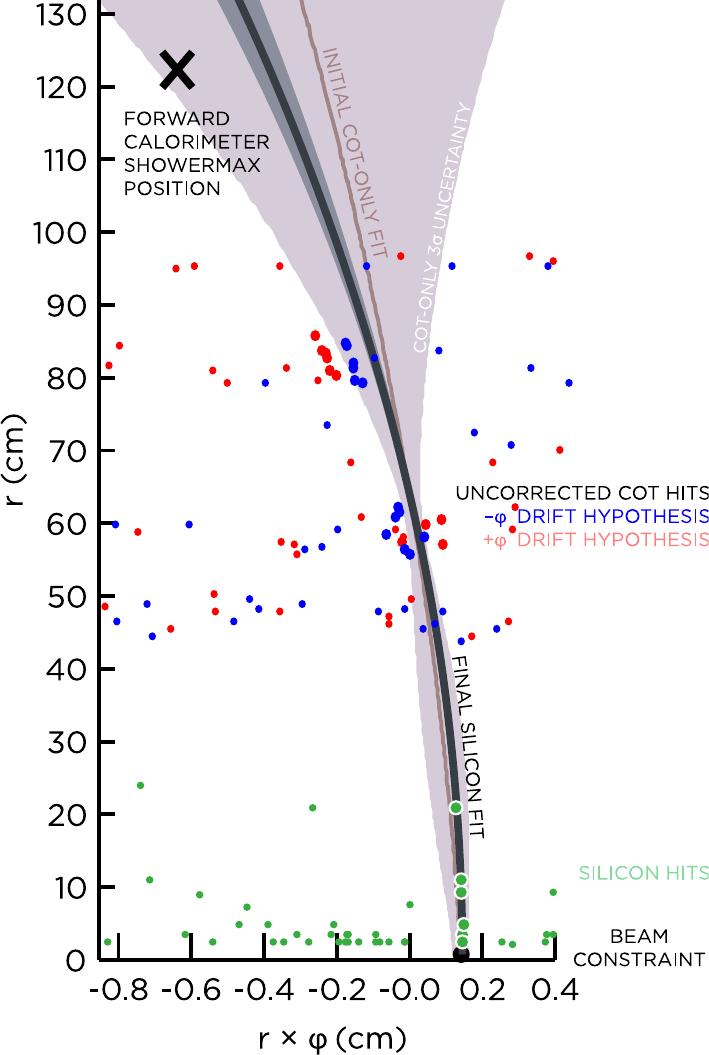}} 
  \caption{A simple example of the Backward tracking algorithm in
  low luminosity data. In a stretched and rotated $r-\phi$ view of the
  relevant section of the tracking volume, COT pulses in a single
  12-layer axial superlayer indicating two possible hit locations
  corresponding to $-\phi$ and $+\phi$ drift are processed to identify
  trajectory segments and fitted to obtain drift time corrections and
  an initial trajectory with large uncertainties. An iterative Kalman
  filter search through possible $\eta$ values for silicon charge
  clusters consistent with the initial trajectory produces a tree of
  track possibilities, from which the single best candidate, shown
  with a projection of the final $3\sigma$ uncertainties, is
  chosen. Also shown is an independent measurement from the forward
  calorimeter shower maximum scintillator.}\label{fig:bw-example}
\end{figure}

After obtaining initial helix fits, including estimates of the helix
uncertainties and correlations obtained from the small number of COT
hits, the algorithm begins an outside-in silicon hit search that uses
a Kalman filter to correct for energy loss and multiple scattering in
the tracker material. After each search completes, quality criteria
based on hit pattern, multiple usage, charge deposition, and module
operational status are applied to the successful hypotheses, and all
surviving hypotheses except those with the most hits are
discarded. Finally, the COT is searched for any remaining information,
including stereo measurements in inner superlayers.

The Backward algorithm has been validated on a variety of samples,
with emphasis on large samples of $\Zboson \rightarrow ee$ and
$\Zboson \rightarrow \mu\mu$ simulation and
data. Fig. \ref{fig:period8-muon-comparison} shows the improvement in
$\Zboson \rightarrow \mu\mu$ yield in muon-triggered data involving
higher-quality forward tracks with COT hits in a subset of the data,
demonstrating the increase in muon acceptance due to the new
software. The lower curve represents the dimuon mass spectrum for the
combination of a trigger muon tracked with the central algorithm and a
forward muon tracked in the combination of the COT and the silicon
detectors with the silicon-driven algorithm in the standard
software. The upper curve shows the same spectrum in the new software,
where the Backward algorithm has largely superseded the other
silicon-driven algorithm. With a modest increase in background, the
peak yield has improved by about 260\%, corresponding to an
approximately 10\% increase in the total \ZMM yield over the entire
detector. The distributions of all forward muon identification
variables are qualitatively the same as those of muons found with the
central COT-driven algorithm, indicating that we have selected a sample of
forward muons with purity comparable to the central muons.

\begin{figure} \centering
  \vskip 0.1in
  \mbox{\includegraphics{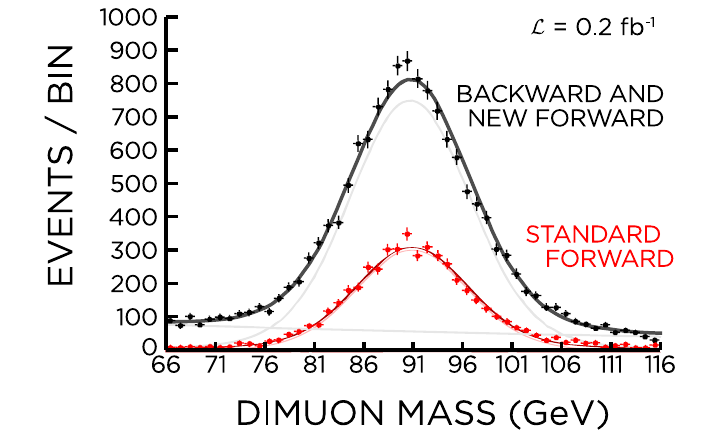} }
  \caption{$\ZMM$ yield and background comparison in $0.2\:\lumfb$ of
    data for candidates combining a trigger muon with another muon
    reconstructed by any dedicated forward tracking algorithm for
    (lower curve) the standard CDF reconstruction and (upper curve) the
    reconstruction used for the present analysis. The gray curves indicate the shapes and normalization of the signal (Breit-Wigner distribution convolved with a Gaussian resolution function) and background (exponential distribution) components used in the fit. In both cases, attached COT hits and the muon identification criteria listed in
    Table \ref{table:muons} are applied.}
    \label{fig:period8-muon-comparison}
\end{figure}
\end{prdappendix}

\bibliographystyle{prd}
\bibliography{prd}{}

\end{document}